\documentclass[journal]{IEEEtran}

\ifCLASSINFOpdf
\else
\fi
\usepackage{cite}
\usepackage{amsmath,epsfig}
\usepackage{algorithm} 
\usepackage{algorithmic, url}
\usepackage{mathtools}
\usepackage{subfigure}
\usepackage{graphicx}
\usepackage{diagbox}

\usepackage{tabularx}
\usepackage{gensymb}
\usepackage{enumitem}
\usepackage{threeparttable}
\usepackage{mathtools}  
\usepackage{amssymb}  
\usepackage{xcolor}
\usepackage{multirow}

\DeclareSymbolFont{matha}{OML}{txmi}{m}{it}
\DeclareMathSymbol{\varv}{\mathord}{matha}{118}

\newcommand*{\Scale}[2][4]{\scalebox{#1}{$#2$}}%
\begin{document}

\title{Sequential Reinforced 360-Degree Video Adaptive Streaming with Cross-user Attentive Network}
%
%
%

\author{Jun~Fu,
        Zhibo~Chen,~\IEEEmembership{Senior Member,~IEEE},
        Xiaoming~Chen, 
        and~Weiping~Li,~\IEEEmembership{Fellow,~IEEE}
\thanks{Jun Fu, Zhibo Chen, and Weiping Li are with the CAS Key Laboratory of Technology in Geo-spatial Information Processing and Application System, University of Science and Technology of China, Hefei 230027, China (e-mail: fujun@mail.ustc.edu.cn; chenzhibo@ustc.edu.cn; wpli@ustc.edu.cn). Xiaoming Chen is a professor from School of Computer Science and Engineering, Beijing Technology and Business University, China (e-mail:xiaoming.chen@btbu.edu.cn). Corresponding Author: Zhibo Chen, Xiaoming Chen.}
\thanks{}
\thanks{}}

%
%

\markboth{IEEE TRANSACTIONS ON Broadcasting SUBMISSION}%
{Shell \MakeLowercase{\textit{et al.}}: Sequential Reinforced 360-Degree Video Adaptive Streaming with Cross-user Attentive Network}
%



\maketitle 


\begin{abstract}
In the tile-based 360-degree video streaming, predicting user's future viewpoints and developing adaptive bitrate (ABR) algorithms are essential for optimizing user's quality of experience (QoE). Traditional single-user based viewpoint prediction methods fail to achieve good performance in long-term prediction, and the recently proposed reinforcement learning (RL) based ABR schemes applied in traditional video streaming can not be directly applied in the tile-based 360-degree video streaming due to the exponential action space. Therefore, we propose a sequential reinforced 360-degree video streaming scheme with cross-user attentive network. Firstly, considering different users may have the similar viewing preference on the same video, we propose a cross-user attentive network (CUAN), boosting the performance of long-term viewpoint prediction by selectively utilizing cross-user information. Secondly, we propose a sequential RL-based (360SRL) ABR approach, transforming action space size of each decision step from exponential to linear via introducing a sequential decision structure. We evaluate the proposed CUAN and 360SRL using trace-driven experiments and experimental results demonstrate that CUAN and 360SRL outperform existing viewpoint prediction and ABR approaches with a noticeable margin.

\end{abstract}


%
\IEEEpeerreviewmaketitle

\section{Introduction}
%
%
%
%
\IEEEPARstart{A}{fter} years of development, Virtual Reality (VR) has reached a new level of technological maturity, and has been increasingly penetrating diverse application areas including entertainment, retail, real-estate, education, healthcare, etc. 360-degree video or panoramic video is a key component of these emerging VR applications. Particularly, 360-degree video allows a user to freely navigate through the captured video scene in a panoramic manner by changing his/her desired viewpoints, offering immersive experience thus significantly enhances presence to users. However, the huge data size of 360-degree video is imposing unprecedented challenges in both storage and transmission. For instance, a premium quality 360-degree video with 120 frames-per-second and 24K resolution can easily consume a bandwidth of multiple Gigabits-per-second (Gbps) \cite{huawei2016}. Also, for smooth rendering, the 360-degree video has to be streamed consistently and reliably at a high rate \cite{huawei2016}. Thus, it is critically significant and imperative to develop effective and efficient methods for 360-degree video transmission. In fact, in 360-degree video, at any given time a user can only watch a portion of the video scene within a field of vision (FoV) centered at a certain direction and with limited horizontal and vertical spans. However, the video service providers, including Youtube \cite{youtube2011youtube}, currently deploy 360-degree video streaming in a straightforward manner that the entire high-quality 360-degree videos are delivered to users. This inevitably leads to a waste of bandwidth.

\begin{figure}[t]
	\centering
	\includegraphics[height=5cm,width=8.5cm]{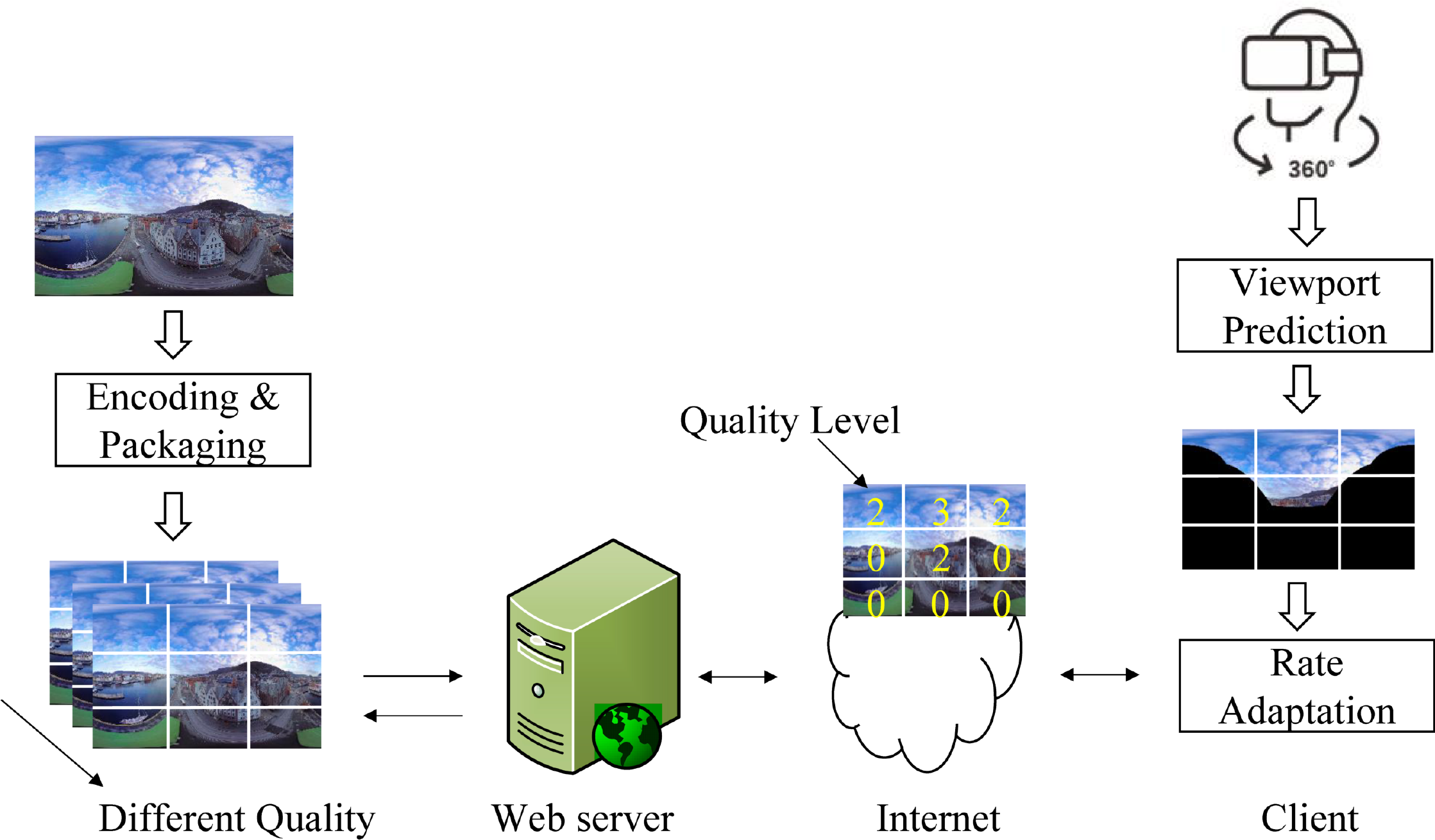}
	\caption {Framework of tile-based 360-degree video streaming.}
	\label{tile-based framework}
\end{figure}

To reduce the transmission bandwidth, many viewpoint-aware panoramic video streaming schemes are proposed based on dynamic adaptive streaming over HTTP (DASH)\cite{stockhammer2011dynamic}, such as asymmetric panoramic streaming \cite{sreedhar2016viewport,xu2018probabilistic}, multi-tier based methods \cite{sun2019two, sun2018multi}, and tile-based panoramic streaming framework \cite{hosseini2016adaptive, xie2017360probdash,ban2018cub360,le2016tiled, xie2018cls, ban2017optimal}. Among them, tile-based panoramic streaming becomes the prevalent approach to transmit 360-degree videos on the internet due to its higher efficiency in storage and transmission. In tile-based streaming, as illustrated in Fig. \ref{tile-based framework}, at the server the 360-degree video is spatially split into several tiles that are independently encoded at various quality levels, while the client conducts viewpoint prediction and rate adaptation to prefetch video segments from the server. However, it is still technically challenging for such a framework to provide users with high quality of experience (QoE) due to the fluctuated network condition and users' irregular head movement. 

Therefore, to deal with the above issues, in this paper we focus on two related research problems, i.e., intelligent viewpoint prediction and reinforcement learning based rate adaptation.

For the first research problem, the existing schemes can be mainly categorized into three categories: 1) Single-user based algorithms \cite{qian2016optimizing, hosseini2016adaptive, xu2018probabilistic, xie2017360probdash}, 2) Content-based algorithms \cite{xu2018gaze, fan2017fixation}, and 3) Cross-user based algorithms \cite{ban2018cub360, xie2018cls}. These schemes, however, suffer from some key limitations: 1) Single-user based algorithms exploit linear regression (LR) model to forecast user's future viewpoints based on user's historical viewpoints, which fails to perform well in long-term prediction due to ignoring the nonlinear characteristics of user's head movement; 2) Content-based approaches utilize motion and saliency \textcolor{black}{maps} of 360-degree videos to improve the performance of viewpoint prediction, since user's viewpoints are largely related to the video content. However, it is time-consuming to extract the motion and saliency \textcolor{black}{maps} from high-definition 360-degree videos. Moreover, the obtained the motion and saliency map  are not reliable enough \cite{xie2018cls}. 3) The competitive cross-user based algorithm, KNN \cite{ban2018cub360}, amends the long-term prediction bias by means of exploiting  $K$  nearest viewpoints  of other  users  around  the  predicted result by the LR model. However, the LR model is easily biased, which inevitably impairs the performance of KNN.

In order to boost the performance of KNN, we aim to design a better viewpoint prediction scheme. Considering users with similar viewing pattern may have similar preference on the future video frames,  we propose a cross-user attentive network called CUAN to predict user's future viewing trajectory. Specifically, we firstly encode user's and cross-user historical viewpoints into hidden vectors via a shared recurrent neural network (RNN), and then conduct viewpoint prediction based on the cross-user information extracted by an attention mechanism.

For the second research problem, the existing methods can be divided into two branches: 1) Combinatorial optimization based schemes \cite{xie2017360probdash,xie2018cls,hosseini2016adaptive} and 2) Reinforcement learning (RL) based schemes \cite{mao2017neural,gadaleta2017d}.  However, there are also several key limitations with these schemes: 1) Combinatorial optimization based methods heavily rely on the accurate estimation of bandwidth budget, e.g., overestimating the bandwidth budget will endanger the playback or even result in rebuffering. Moreover, these methods typically do not take the cascading effect of birtate decision into consideration; 2) Reinforcement learning (RL) based approaches have shown its promising potential in tackling with fluctuated network condition and various QoE objectives. However, these schemes can not be directly applied in tile-based 360-degree video streaming owing to the exponential action space (e.g., given N tiles and each tile has M bitrate level, the dimension of action space is $M^N$). 

Therefore, we propose a sequential RL-based method, called 360SRL. Specifically, we introduce a sequential decision structure, i.e., selecting bitrate for each tile in sequence instead of determining the bitrate of all tiles in one step, which transforms the action space size of each decision step from exponential to linear. Then, 360SRL learns to make ABR decisions solely through thousands of interactions with the deployment environment, instead of relying on pre-programmed models or assumptions about the deployment environment. As a result, 360SRL is able to learn ABR algorithms that adapt to a wide range of environments and QoE objectives. 

The main contributions of this paper can be summarized as follows: 
\begin{itemize}
  \item We propose a cross-user attentive network for viewpoint prediction, named CUAN, which boosts the performance of viewpoint prediction through exploiting cross-user information extracted by an attention mechanism. 
  
 \item We propose a sequential RL-based method for rate adaptation, called 360SRL, which successfully applies RL in the tile-based 360-degree video streaming via introducing a sequential decision structure.
 
 \item We integrate CUAN and 360SRL into a 360-degree video adaptive streaming framework. The experimental results show that our prototype outperforms existing methods with a noticeable margin.
\end{itemize}

\textcolor{black}{This journal paper is an extension of our published short conference paper \cite{fu2019360srl}, the key differences lie in three aspects. First, this journal version sets up a complete panoramic video streaming framework with more theoretical analysis, and solves another challenging problem in panoramic video streaming, i.e., viewpoint prediction with a novel cross-user attentive network. Second, this journal version introduces a simple but efficient cross-user attention mechanism to mitigate error in viewpoint prediction, which is validated with detailed experimental analysis. Third, more thoroughly experiment results and ablation studies are provided in this paper to verify effectiveness of the proposed framework.}

The remainder of this paper is organized as follows. Section 2 discusses the related work concerning tile-based 360-degree video streaming. Section 3 details the proposed approach for viewpoint prediction and rate adaptation in sequence. Then, performance evaluation and comparison are presented in Section 4. Finally, Section 5 concludes the paper and discusses some future research directions.

\section{Related Work}

\subsection{Viewpoint Prediction}
For viewpoint prediction, the existing algorithms can be categorized into three classes: single-user based ones, content based ones  and cross-user based ones. In single-user based category, Qian et al. \cite{qian2016optimizing} and Stefano et al. \cite{petrangeli2017http} employ a Linear Regression (LR) model to forecast user's future viewpoints. To boost the performance of the LR model, Lan et al. \cite{xie2017360probdash} and Xu et al. \cite{xu2018probabilistic} put forward a probabilistic model estimating the distribution of LR's prediction errors. However, the long-term prediction bias is still large, since the assumption of linear head movement is easily violated. In content-based category, Fan et al. \cite{fan2017fixation} and Xu et al. \cite{xu2018gaze} leverage saliency and motion maps of 360-degree videos and RNN to conduct viewpoint prediction. However, extracting saliency and motion maps from high-definition 360-degree videos \textcolor{black}{is} high-computation. Besides, the generated saliency and motion maps are not reliable enough \cite{xie2018cls}. In cross-user based category, CLS \cite{xie2018cls} groups others' viewpoints into clusters via a density-based clustering algorithm \cite{ester1996density}, and then \textcolor{black}{uses} SVM \cite{chang2011libsvm} to predict the cluster to which the current user \textcolor{black}{belongs} according to his/her past viewing trajectory. KNN \cite{ban2018cub360} \textcolor{black}{amends} the prediction bias of the LR model through utilizing $K$ nearest viewpoints of other users around the predicted result. However, the long-term prediction bias still exists in KNN, since KNN is based on LR. Thus, in this paper, we aim to develop a better viewpoint prediction algorithm to improve the performance of KNN.

\subsection{Rate Adaptation}
The existing rate adaptation approaches could be divided into two branches: combinatorial optimization based \textcolor{black}{methods} and reinforcement learning (RL) based \textcolor{black}{methods}. In combinatorial optimization based category, these schemes typically formulate rate adaptation as a variant of multiple-choice knapsack (MCKP) problem \cite{sinha1979multiple}, which aim to maximize the defined QoE objective given a bandwidth budget. The available bandwidth estimation methods include rate-based ones \cite{jiang2014improving,sun2016cs2p}, buffer-based ones \cite{huang2014buffer,spiteri2016bola}, and target-buffer based ones \cite{xie2017360probdash}. Since the rate adaptation is a NP-hard optimization problem, brute-forced search seems infeasible. As a result, 360ProDash \cite{xie2017360probdash}, CLS \cite{xie2018cls}, and  Hosseini et al.  \cite{hosseini2016adaptive} present a greedy algorithm, dynamic programming algorithm, and divide-and-conquer approach to solve the MCKP problem at a relatively low-complexity manner, respectively. However, these schemes fail to achieve optimal performance across a wide variety of network conditions since the bandwidth budget is estimated via fixed rules on simplified or inaccurate models of the deployment environment \cite{mao2017neural}. Besides, these methods do not take the cascading effect of bitrate decision into consideration, e.g., CLS \cite{xie2018cls} aims to maximize the quality of user's FoV regardless of the change of bitrate between two consecutive FoVs. In RL-based category, Pensieve \cite{mao2017neural} and D-Dash \cite{gadaleta2017d} have  shown their promising potential of dealing with various network condition and QoE metrics in traditional video streaming. However, these \textcolor{black}{methods} can not be directly applied in tile-based streaming system due to the exponential action space. To reduce action space size, the recent method, DRL360 \cite{zhang2019drl360}, allocates the same bitrate for tiles within users' FoV regardless of their different importance, which inevitably limits user's perceived video quality.

\section{Proposed Method}
\begin{figure*}[t]
	\centering
	\includegraphics[width=1\linewidth]{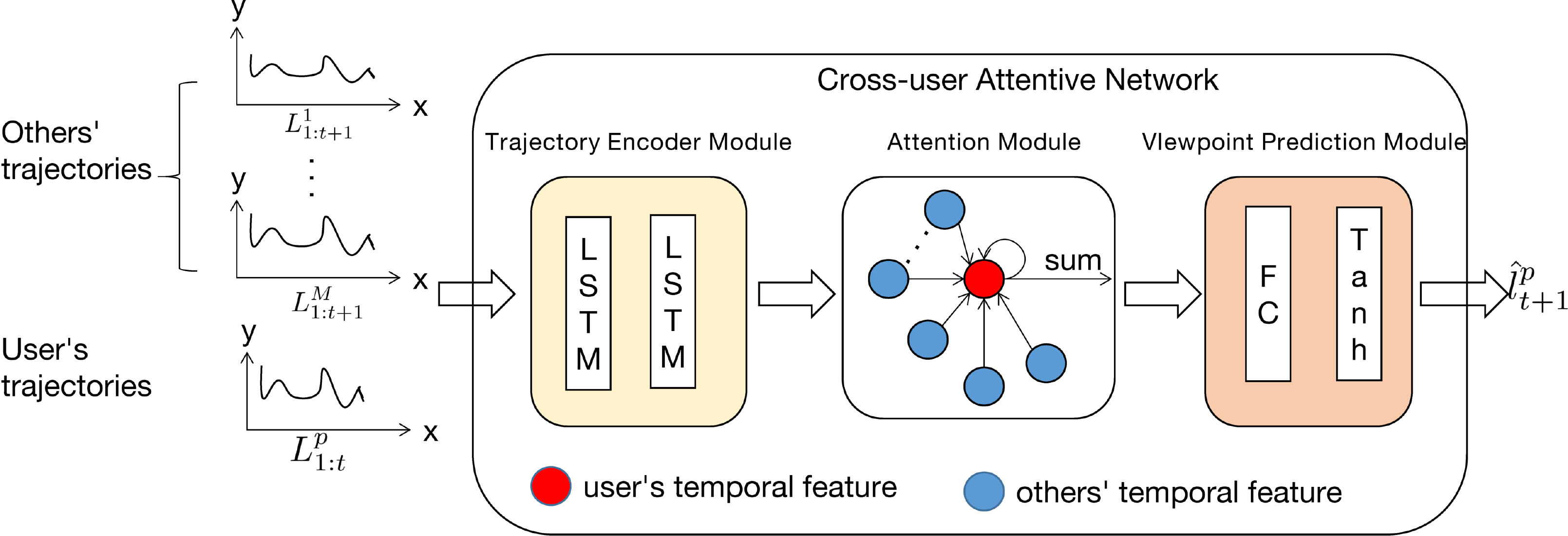}
	\caption {\textcolor{black}{The illustration of the proposed cross-user attentive network, comprised of a trajectory encoder module, an attention module, and a viewpoint prediction module. As shown, the trajectory encoder module is made up of two stacked Long Short-Term Memory (LSTM) layers, the attention module adopts the self-attention mechanism, and the viewpoint prediction module consists of one fully-connected (FC) layer and an activation function (Tanh). }}
	\label{gcn}
\end{figure*}
\subsection{Viewpoint Prediction}
\subsubsection{Problem Formulation}
We formulate the viewpoint prediction in the tile-based 360-degree video streaming as follows: Given historical viewpoints of the $p^{\text{th}}$ user $L^p_{1:t} =\{{l}^p_1, {l}^p_2,...,{l}^p_t\}$ where ${l}^p_t = (x_t,y_t)$, $x_t \in [-180,180]$ and $y_t \in [-90, 90]$ correspond to the longitude and latitude of the viewpoint on a 3D sphere, and other users' viewpoints on the same video $L^{1:M}_{1:t+T}$ where $M$ denotes the number of other users, then viewpoint prediction aims to forecast the future $T$ viewpoints: ${L}^p_i$ where $i = t+1,...,t+T$. Mathematically, the objective of viewpoint tracking can be expressed as follows:
\begin{equation}
	\mathop{\min}_{F} \ \ \sum_{k=t+1}^{t+T} \|l^p_k-\hat{l}^p_k\|_1,
\end{equation}
where $F$ denotes the viewpoint prediction model and $\hat{l}^p_k$ represents the predicted result by the $F$. 

The existing competitive cross-user based approach, KNN \cite{ban2018cub360}, chooses the LR model to model $F$ and reduces the long-term prediction bias via exploiting K nearest viewpoints of other users around the predicted result. However, the LR model is easily biased, which inevitably impairs the performance of KNN. As a result, to boost the performance of KNN, we propose a cross-user attentive network to model $F$, which takes user's historical viewing trajectory and cross-user viewpoints into consideration. The motivation of this design is two-fold. On the one hand, user's historical viewpoints provide key clues for viewpoint prediction, since they reveal user's unique viewing pattern in exploring a video scene. On the other hand, considering users with similar viewing pattern may have similar preference on the future video frames, we propose \textcolor{black}{an} attention mechanism to automatically extract useful cross-user information from viewpoints of other users. \textcolor{black}{Although the attention mechanism has been proven to be effective in various tasks, such a technique needs to be carefully designed to accommodate specific problems. In the task of viewpoint prediction, it is intuitive to pay more attention to the viewpoints of other users who have similar preference on the same video as the current user when generating representations of cross-user information. In this paper, the similarity between other users and the current user is calculated based on the past viewing trajectories of other users and the current user, rather than the single-timestamp viewpoints of other users and the current user \cite{li2019very}. Intuitively, such a method of calculating similarity is more reasonable than the previous method \cite{li2019very}, which is verified by the experimental results.}


\subsubsection{Cross-user Attentive Network}
As shown in Fig. \ref{gcn}, our proposed method consists of a trajectory encoder module, an attention module, and a viewpoint prediction module. Next, we will detail these modules in sequence. 

The trajectory encoder module aims to extract temporal features from users' historical viewpoints. Inspired by the good performance of Recurrent Neural Network (RNN) \cite{cleeremans1989finite} for capturing the sequential information \cite{xu2018gaze}, we employ the Long Short-Term Memory (LSTM) \cite{gers2000recurrent}, a variant of RNN, to encode the viewing path of a user. Specifically, for predicting the viewpoint at the $(t+1)^{th}$ frame, we firstly feed the historical viewpoints of the $p^{th}$ user into the LSTM as follows:
 \begin{equation}
     f^p_{t+1} = h({l}^p_1, {l}^p_2,...,{l}^p_t),
 \end{equation}
 where the function $h(\cdot)$ denotes the input-output function of the LSTM. Then, we use the same LSTM to encode others' viewing trajectories as follows:
 \begin{equation}
 f^i_{t+1} = h({l}^i_1, {l}^i_2,...,{l}^i_{t+1}), i \in \{1,.., M\}.
\end{equation}

The attention module is designed to extract information related to the $p^{th}$ user from others' viewpoints. Firstly, we derive the correlation coefficient of the $p^{th}$ user and others as follows:
\begin{equation}
 s^{pi}_{t+1} = z({f}^i_{t+1}, {l}^p_{t+1}), i \in \{1,.., M\} \cup \{p\},
\end{equation}
where $s^{pi}_{t+1}$ represents the similarity of the $p^{th}$ user and the $i^{th}$ user, and the function $z$ is modeled by inner product. It is worth noting that $z$ can be modeled by other ways, such as \textcolor{black}{multiple fully-connected (FC) layers}. Then, we normalize the correlation coefficient as follows:
\begin{equation}
 \alpha^{pi}_{t+1} = \frac{e^{s^{pi}_{t+1}}}{\sum\limits_{i \in \{1,..., M\} \cup \{p\}} e^{s^{pi}_{t+1}} },
\end{equation}
Finally, we obtain the fused feature as follows:
\begin{equation}
 g^{p}_{t+1} = {\sum\limits_{i \in \{1,...,M\} \cup \{p\}} \alpha^{pi}_{t+1} \cdot f^i_{t+1}}.
\end{equation}

The viewpoint prediction module aims to forecast the future viewpoints according to the fused feature generated by the attention module. Specifically, the viewpoint at $(t+1)^{th}$ frames can be estimated as follows:
\begin{equation}
    \hat{l}^p_{t+1} = r(g^{p}_{t+1}),
\end{equation}
where the function $r(\cdot)$ is modeled by \textcolor{black}{one FC layer}. It is worth noting that  viewpoints corresponding to future T frames are predicted in a rolling fashion. 

\subsubsection{Loss function} 
The loss function is defined as the sum of the all the absolute differences between the predict viewpoints and the ground truth, which can be formulated as follows:
\begin{equation}
    L = \sum_{i=t}^{t+T}  \lvert\lvert \hat{l}^p_{i} - l^p_{i} \rvert\rvert_1.
\end{equation}

\subsubsection{Implement Details}
We implement the proposed CUAN on the popular deep-learning framework, PyTorch \cite{paszke2017pytorch}. The function $h(\cdot)$ \textcolor{black}{is made up of two stacked LSTM layers}, both with 32 neurons. The function $r(\cdot)$ consists of \textcolor{black}{one FC layer  with 32 neurons}, followed by a Tanh function. The length of historical viewpoints and future viewpoints are set to 1 second and 5 seconds. During training, 2048 samples are randomly generated from the dataset per iteration. All trainable variables of the proposed CUAN are optimized by Adam \cite{kingma2014adam}, where $\beta_1$, $\beta_2$ and $\epsilon$  are set to 0.9, 0.999, and $10^{-8}$, respectively. The learning rate and the number of training epoch are set to $10^{-3}$ and 50.

\subsection{Rate Adaptation}

\subsubsection{Problem Formulation}
In the tile-based 360-degree video streaming, a 360-degree video is cut to $m$ $T$-second video segments, and each video segment is spatially split into $N$ tiles that are independently encoded at $M$ bitrate levels. Hence, there are $N \times M$ optional encoded chunks for each segment. Rate adaptation aims to find the optimal bitrate set $X=\{x_{i,j}\} \in Z^{N \times M}$ (where $x_{i,j} = 1$ means choosing $j^\text{th}$ bitrate level for $i^\text{th}$ tile and $ x_{i,j} = 0 $ otherwise) for each segment to maximize user's QoE objective. Mathematically, this problem can be formulated as follows:
\begin{equation}
	\mathop{\max}_{X} \ \ \sum_{t=1}^{m} Q_t,
\end{equation}
where $Q_t$ denotes the QoE score of the $t^\text{th}$ segment.

In this paper, referring to existing QoE models \cite{qian2016qoe,yu2017qoe,ban2018cub360,zhang2019drl360,doumanoglou2018quality,tam2011stereoscopic}, the QoE score of a certain segment is related to the following aspects:
\begin{itemize}
    \item Viewport Quality [Mbps]: As only the video content within user's FoV is rendered, the video quality merely depends on the viewport quality, which can be calculated as follows:
    \begin{equation}
    	 Q^1_t  = \sum_{i=1}^N \sum_{j=1}^{M} x_{i,j} \cdot p_i \cdot r_{i,j},
    \end{equation}
    where $p_i$ denotes the normalized viewing probability of the $i^\text{th}$ tile and $r_{i,j}$ records the bitrate of chunk $(i,j)$. It is worth noting that we suppress the \textcolor{black}{segment} index in $x_{i,j}$, $p_i$, and $r_{i,j}$ for compactness and clarity.
    
    \item Viewport Temporal Variation [Mbps]: Since drastic change of rate between two consecutive viewports may cause users dizziness and headache \cite{zhang2019drl360}, viewport temporal variation is also taken into consideration, which can be measured via: 
     \begin{equation}
    	Q^2_t = \lvert Q^1_t - Q^1_{t-1} \rvert.
    \end{equation}     

    \item Viewport Spatial Variation [Mbps]: We also take the viewport spatial variation into account because rate changes among tiles within user's FoV may introduce blocking artifacts.  The specific expression of the viewport spatial variation is written as follows:
    \begin{equation}
    	Q^3_t = \frac{1}{2} \sum_{i=1}^{N} \sum\limits_{u \in U_i} p_i \cdot p_u  \sum_{j=1}^M \lvert x_{i,j} \cdot r_{i,j}-x_{u,j} \cdot r_{u,j} \rvert,
    \end{equation}
    where $U_i$ denotes the index of tiles in the 1-hop neighbor of the $i^{th}$ tile \cite{ban2018cub360}. 
    
    \item Rebuffering [s]: Rebuffering typically occurs when the downloading time of a segment is greater than the buffer occupancy of the client player, which is the most annoying to video-streaming users \cite{chen2015multimedia}. Thus,  rebuffering is also a key aspect of user's QoE \cite{de2012quantifying}, which can be derived by:
   \begin{equation}
	Q^4_t = \text{max}( \frac{\sum_{i=1}^N \sum_{j=1}^{M} x_{i,j} \cdot r_{i,j} \cdot T}{\xi_t} - b_{t-1}, 0),
    \end{equation} 
    where $\xi_t$ and $b_{t-1}$ denote the network throughput and the buffer occupancy of the video player.
\end{itemize}
Based on the aforementioned analysis, the QoE objective is defined as follows:
\begin{equation}
	Q_t = Q^1_t - \eta_1 \cdot Q^2_t - \eta_2 \cdot Q^3_t - \eta_3 \cdot Q^4_t,
	\label{hello}
\end{equation}
where $\eta_*$ are adjustable parameters and different $\eta_*$ means different user's preference.

\begin{figure}[t]
	\centering
	\includegraphics[height=3.5cm,width=9cm]{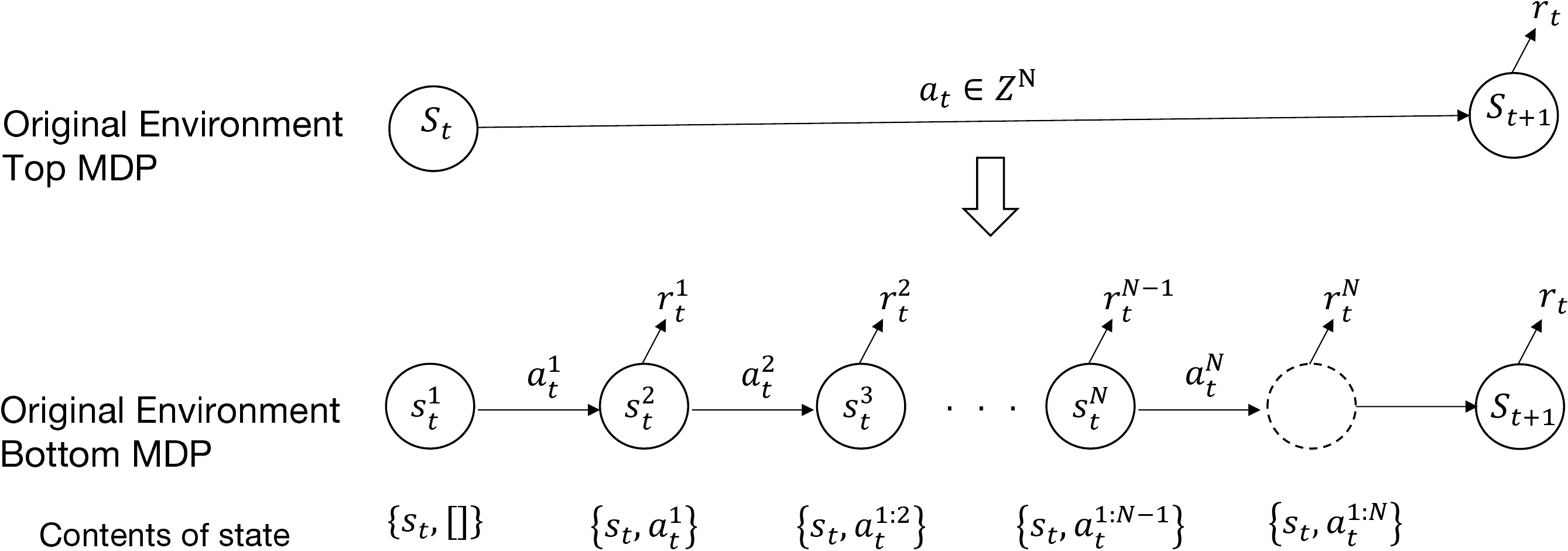}
	\caption {Illustration of Sequential Decision Structure.}
	\label{framework}
\end{figure}

\subsubsection{Sequential RL-based ABR Scheme} Similar to ordinary video streaming \cite{mao2017neural,gadaleta2017d}, we assume that the ABR decision process in tile-based 360-degree video streaming can be also regarded as a Markov Decision Process (MDP) illustrated by the ``Top MDP'' in Fig. \ref{framework}. The ``Top MDP'' shows that, at a specific time, e.g. $t$, an agent performs a $N$-dim action $a_t \in Z^N$ (the bitrate set of the $t^\text{th}$ segment) derived from the observed state $s_t$ to the deployment environment, and then receives a feedback signal $r_t$ that reflects user's QoE. It can be observed that the “Top MDP” has $M^N$ possible actions at each decision step, which hinders the application of the RL-based ABR algorithms \cite{mao2017neural,gadaleta2017d} due to the exponential action space. To resolve this problem, we propose a novel sequential decision structure to reduce the action space of one step from $M^N$ to $M$. Specifically, we stretch out one transition with a $N$-dim action in the ``Top MDP'' into $N$ cascading transitions with a 1-dim action in the ``Bottom MDP'', which are tied together with Bellman Equation \cite{singh1996make}. As illustrated in the ``Bottom MDP'', the agent derives a 1-dim action $a^i_t$ from the state  $s^i_t$ composed of the original state $s_t$ and the set of previous selected actions $\{a^1_t,...,a^{i-1}_t\}$, then gets the feedback signal $r^i_t$ recording the value of $a^i_t$. At last, the agent receives \textcolor{black}{an reward} $r_t$ indicating the value of the action $a_t$. It is worth noting that, according to our experimental results, the decision order of tiles in the ``Bottom MDP'' is set as follows: the agent firstly picks up the lowest bitrate for tiles out of user's FoV, and then makes bitrate decision for tiles within user's FoV in the order of viewing probabilities from high to low. Next, we will introduce the design of the agent in detail. 

The state, action and reward representation of our agent are constructed as follows:
\begin{itemize}
	\item State. \textcolor{black}{The state describes the situation when we prefetch a new segment, typically including the network bandwidth, the buffer size of the player, and the file size of the upcoming segment. Specifically,} for the $i^\text{th}$ tile in the $t^\text{th}$ segment, the state is defined as follows:
	\begin{equation}
	    s^i_t = \{\vec{\tau}_{t,i}, \vec{P}_{t,i}, \vec{q}_{t,i}, \hat{\xi}_t, Q^1_{t-1},  p^i_t,  b^i_{t-1}\},
    \end{equation}
	 where  $\vec{\tau}_{t,i}$ is a $M$-dim vector recording the file size of $i^\text{th}$ tile at different bitrate levels; $\vec{P}_{t,i}$ and $\vec{q}_{t,i}$  are the viewing probabilities and chosen bitrate of tiles within 1-hop neighbor of the $i^\text{th}$ tile; $\hat{\xi}_t$ is network throughput estimated by the harmonic mean of the experienced throughput for the past five segment downloads;  $Q^1_{k-1}$ records the viewport quality of the last segment; $p^i_t$ denotes the predicted viewing probability of the $i^\text{th}$ tile; $b^i_t$ is the estimated buffer occupancy of the video player calculated by the following equation: $b^i_t=b_{t-1}-\frac{\sum_{h=1}^{i-1} r_{h,a^h_t} T}{\hat{\xi}_t}$.
	 
	\item Action. The output of our agent is an $M$-dim vector, where each entry corresponds to the probability of choosing a certain  bitrate based on the current state $s^i_t$. The entry with the highest probability is chosen as $a^i_t$.

	\item Reward. Reward assesses the value of chosen action being in a certain  state, which guides the agent to learn ABR algorithm. In this paper, the reward $r_t$ in the ``Top MDP'' is measured by the Eq. \ref{hello}. However, in the ``Bottom MDP'', the reward of single transition $r^i_t$ is not available because the original environment does not make changes until a decision is made for the last transition. Hence, we modify the Eq. \ref{hello} to estimate $r^i_t$ as follows:
    \begin{align}
 	    r^i_t = Q^1_{t,i} - \eta_1 \cdot Q^2_{t,i} - \eta_2 \cdot Q^3_{t,i} - \eta_3 \cdot Q^4_{t,i}, \\
	    Q^1_{t,i} = p^i_t \cdot r_{i,a^i_t}, \\ 
	    Q^2_{t,i} = p^i_t \cdot \lvert r_{i,a^i_t} - Q^1_{t-1} \rvert, \\
	    Q^3_{t,i} = \frac{1}{2} \cdot \sum\limits_{u \in U_i} \delta^u_t \cdot p^i_t \cdot p^u_t \sum_{j=1}^M \lvert  r_{i,a^i_t}-x_{u,j} \cdot r_{u,j} \rvert,	    \\
	     Q^4_{t,i} = \text{max}(\frac{r_{i,a^i_t} * T}{\hat{\xi}_t} - b^{i}_t, 0),
    \end{align}
 where $\delta^u_t=0$ means the birate of the $u^\text{th}$ tile is not determined and $\delta^u_t=1$ otherwise. 
\end{itemize}

%
\begin{algorithm}[tb]
	\caption{Training Methodology of 360SRL}
	\begin{algorithmic}[1]
    \REQUIRE 
	  discounting factor in the ``Top MDP'', $\gamma_1$; discounting factor in the ``Bottom MDP'', $\gamma_2$; global shared parameter vectors, $\theta$, $\theta_v$; global shared counter $C$; thread-specific parameter vectors, $\theta'$, $\theta'_v$; maximum number of iterations, $C_{max}$.
	\ENSURE global shared parameter vectors, $\theta$, $\theta_v$.
    \REPEAT
	\STATE Initialize the thread step counter $t \leftarrow 1$
    \STATE Reset gradients: $d\theta \leftarrow 0$ and  $d\theta_v \leftarrow 0$
    \STATE Synchronize thread-specific parameter vectors: $d\theta \leftarrow \theta'$ and  $d\theta_v \leftarrow \theta'_v$
	    \STATE Reset the deployment environment and get state $s_t$
        \REPEAT
            \FOR{$i \in \{1,...,N\}$ }
                \STATE Obtain $a^i_t$ according to the policy $\pi(a_t|s^i_t;\theta'_v)$
                \STATE Calculate the single-step reward $r^i_t$ 
            \ENDFOR
            \STATE Perform $a_t \in Z^N$ to the deployment environment
            \STATE Receive reward $r_t$ and new state $s_{t+1}$
            \STATE $C \leftarrow C + 1$
            \STATE $t \leftarrow t + 1$
        \UNTIL{download all segments of a 360-degree video}
        \STATE $R \leftarrow 0$
     	\FOR{$j \in \{t-1,...,1\}$ }
            \STATE  $ R \leftarrow r_j + \gamma_1 R$
            \STATE  $ R' \leftarrow 0 $            
            \FOR{$i \in \{N,...,1\}$}
               \STATE  $ R' \leftarrow r^i_j + \gamma_2 R' $
               \STATE  $ r \leftarrow R + R' $   
               \STATE Accumulate gradients w.r.t $\theta'$: \\ $d\theta \leftarrow d\theta + \nabla_{\theta'} \text{log} \pi_{\theta'}(s^i_j, a^i_j) (r - V(s^i_j;\theta'_v)) + \beta \nabla_{\theta'} H(\pi(s^i_j;\theta'))$
               \STATE Accumulate gradients w.r.t $\theta'_v$: \\ $d\theta_v \leftarrow d\theta_v + \partial(r - V(s^i_j;\theta'_v))^2 / \partial \theta'_v $               
            \ENDFOR
    	\ENDFOR 
    	\STATE Perform asynchronous update of $\theta$ using $d\theta$ and of $\theta_v$ using $d\theta_v$
    \UNTIL{$C > C_{max}$}
	\end{algorithmic}
	\label{alg: 1}
\end{algorithm}

\subsubsection{Training Methodology}
In this paper, we use the state-of-the-art actor-critic RL technique A3C \cite{mnih2016asynchronous} as the fundamental training algorithm of our agent. In the framework of A3C, multiple agents are training asynchronously and each agent consists of a policy network and a value network. The policy network aims at adjusting its parameters in the direction of increasing the accumulated discounted reward. The gradient of the cumulative discounted reward with respect to the policy parameters, $\theta$, can be written as: 

\begin{equation}
\nabla_\theta E_{\pi_\theta} \lbrack \sum_{t=0}^{\infty} \gamma^t r_t \rbrack =  E_{\pi_\theta}\lbrack \nabla_\theta \text{log} \pi_\theta(s, a) A^{\pi_\theta}(s,a) \rbrack,
\end{equation}
where $ A^{\pi_\theta}(s, a)$ is called the advantage function, which indicates how much better a specific action is compared to the “average action” taken according to the policy. In practice, the unbiased estimate of $A^{\pi_\theta}(s_t, a_t)$ is replaced with the empirically computed advantage $A(s_t, a_t)$ \cite{sutton1998reinforcement}. Therefore, the gradient of policy network can be rewritten as:
\begin{equation}
\theta \leftarrow \theta + \alpha \sum_{t} \nabla_\theta \text{log} \pi_\theta(s_t, a_t) A(s_t,a_t),
\label{grd}
\end{equation}
where $\alpha$ is the learning rate. The advantage $A(s_t,a_t)$ satisfies the following equality: $A(s_t, a_t)=Q(s_t,a_t)-V^{\pi_\theta}(s_t)$, where $V^{\pi_\theta}(s_t)$ represents the value of the current state $s_t$ and $Q(a_t ,s_t)$ is the value of the state-action pair $(s_t, a_t)$. To calculate the advantage $A(s_t, a_t)$, we use the value network to learn the value function of state $V^{\pi_\theta}(s_t)$  from empirically observed rewards $r_t$. Following the standard Temporal Difference method \cite{sutton1998reinforcement},  we update parameters of the value network, $\theta_v$, as follow:
\begin{equation}
\Scale[0.9]{
	\theta_v \leftarrow \theta_v - \alpha_{v} \sum_{t} \nabla_{\theta_v}(r_t+\gamma V^{\pi_\theta}(s_{t+1};\theta_v)- V^{\pi_\theta}(s_{t};\theta_v))^{2}},
\end{equation}
where $\alpha_{v}$ is the learning rate of the value network and $\gamma$ is the discounted factor. Finally, to ensure that the agent discovers good policies, we add an entropy regularization term to Eq. \ref{grd} as follows: 
\begin{equation}
\Scale[0.9]{\theta \leftarrow \theta + \alpha \sum_{t} \nabla_\theta \text{log} \pi_\theta(s, a) A^{\pi_\theta}(s,a) + \beta \nabla_\theta H(\pi_{\theta}(\cdot |s_t))},
\end{equation}
where $H(\cdot)$ is the entropy of the policy at each time step and $\beta$ represents the strength of the entropy regularization term. The detail of training methodology is presented in Algorithm. \ref{alg: 1}.

\subsubsection{Implement details}
\begin{figure}[t]
	\centering
	\includegraphics[width=1\linewidth]{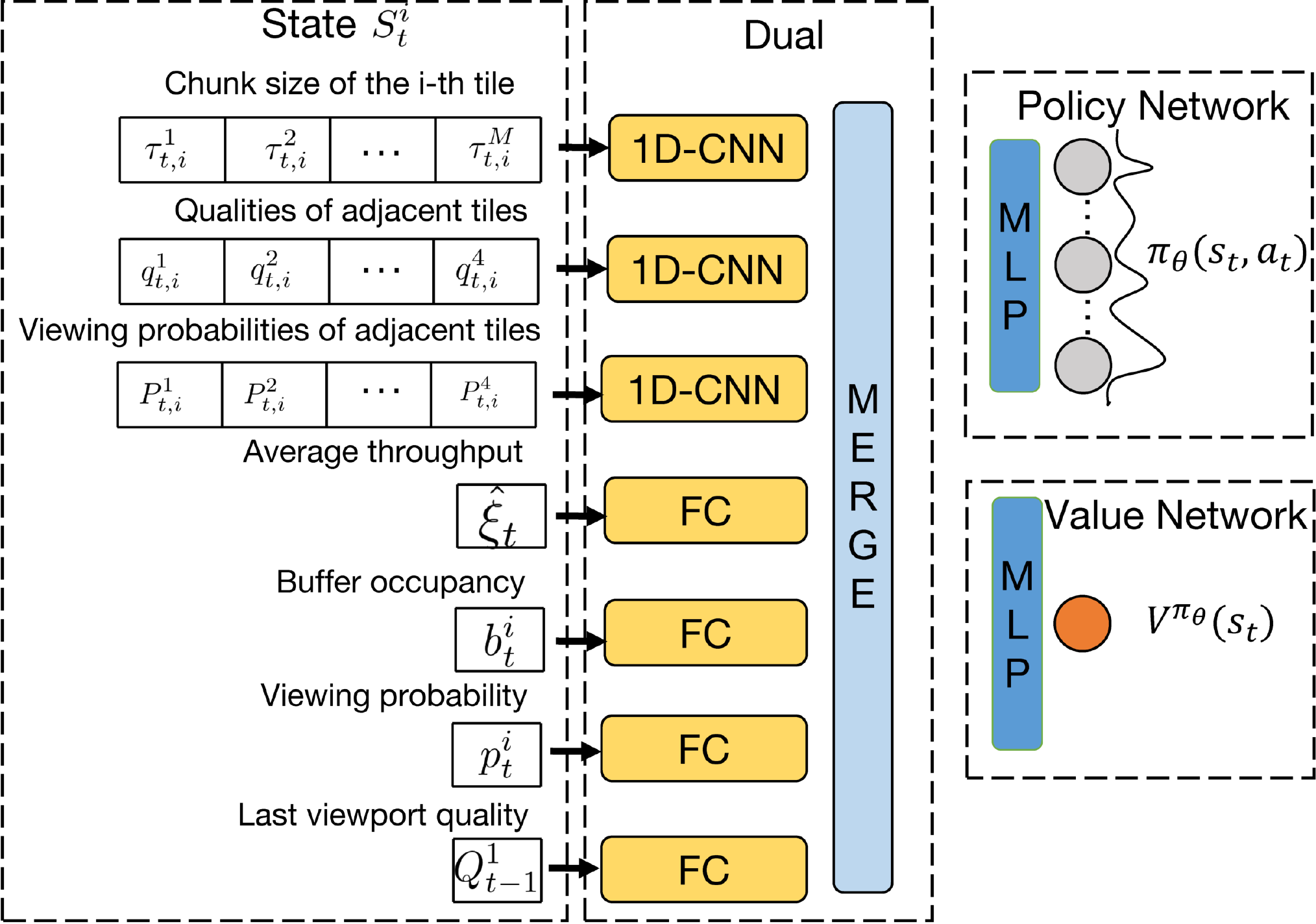}
	\caption{\textcolor{black}{The diagram of the proposed 360SRL, comprised of a policy network and a value network.  As shown, the two networks extract features from the input state via 1D convolutional neural networks (1D-CNNs) and fully-connected (FC) layers, then concatenate the resulting features via the merge block, finally generate a distribution over available bitrate levels and a scalar through multi-layer perceptron (MLP).}}
	\label{fig:RL}
\end{figure}
We use PyTorch \cite{paszke2017pytorch} to implement our agent, too. As demonstrated in Fig. \ref{fig:RL}, our agent feeds the chunk size of the $i^\text{th}$ \textcolor{black}{tile}, the viewing probabilities and chosen bitrate of tiles within the 1-hop neighbor of the $i^\text{th}$ tile into 1D convolutional network (1D-CNN) layers with 64 filters, each of size 3 with stride 1. Results from these layers are then aggregated with other inputs in \textcolor{black}{a multi-layer perceptron (MLP) comprised of one FC layer with 64 neurons} to apply the softmax function. The value network uses the same network architecture, but its final output is a single neuron without activation function. During training, we set the discount factor of ``Top MDP'' and ``Bottom MDP'' to 0.99 and 1, use 16 parallel asynchronous agents for exploration, and set the number of iterations to $10^7$. The parameters of each agent is optimized by Adam \cite{kingma2014adam},  where $\beta_1$, $\beta_2$ and $\epsilon$ are set to 0.9, 0.999, and $10^{-8}$, respectively. The learning rates for the policy and value network is set to $10^{-4}$ and $10^{-3}$. Besides, we initialize the discounted factor $\beta$ as 1, and use a step decay schedule to drop $\beta$ by 0.1 every $10^6$ steps. 
\section{Experiment}
\subsection{Viewpoint Prediction}
\begin{figure*}[t]
    \subfigure{
        \begin{minipage}[t]{0.33\linewidth}
        \centering
        \includegraphics[width=1\linewidth]{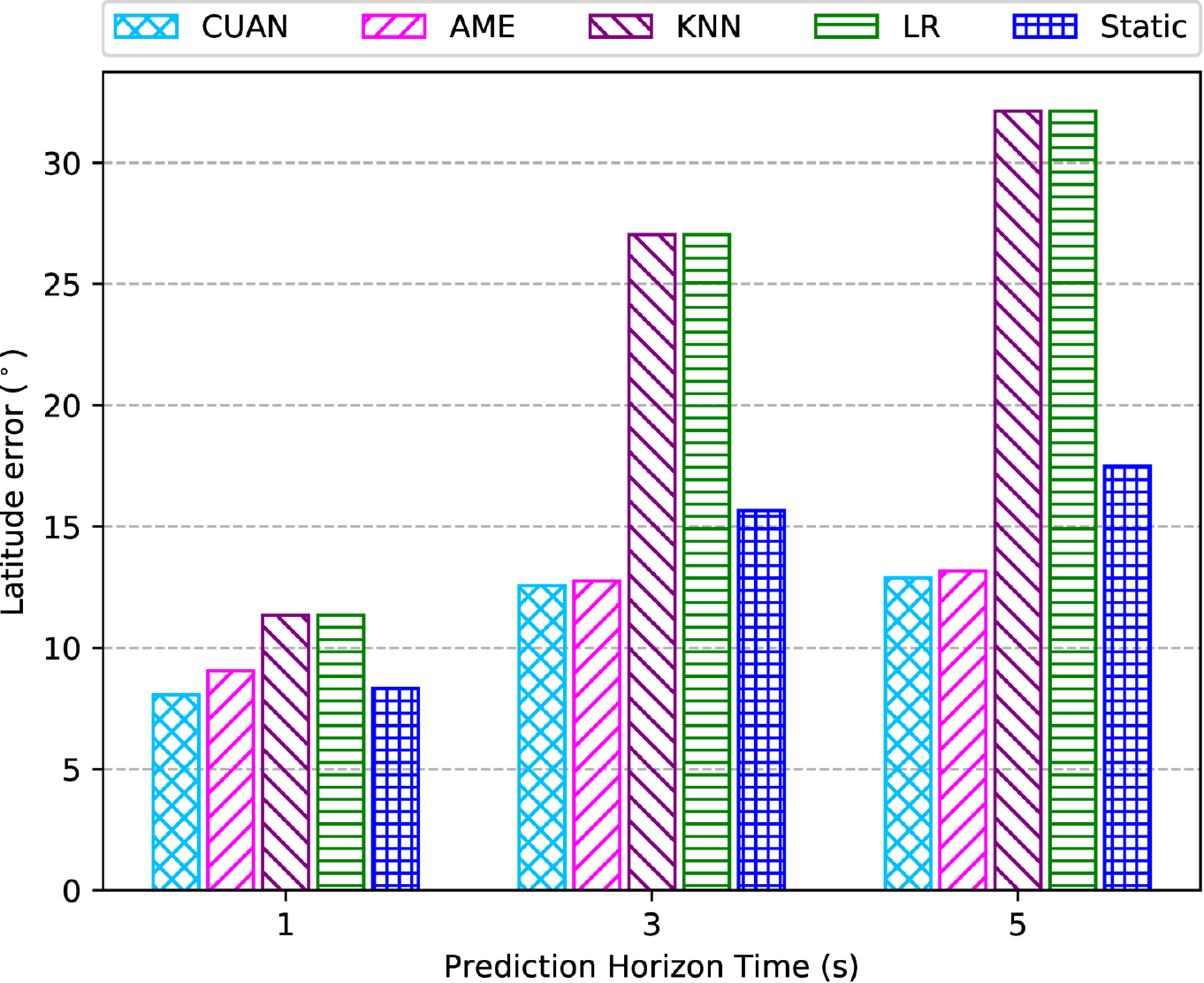}
    \end{minipage}%
    }%
    \subfigure{
	\begin{minipage}[t]{0.33\linewidth}
		\centering
		\includegraphics[width=1\linewidth]{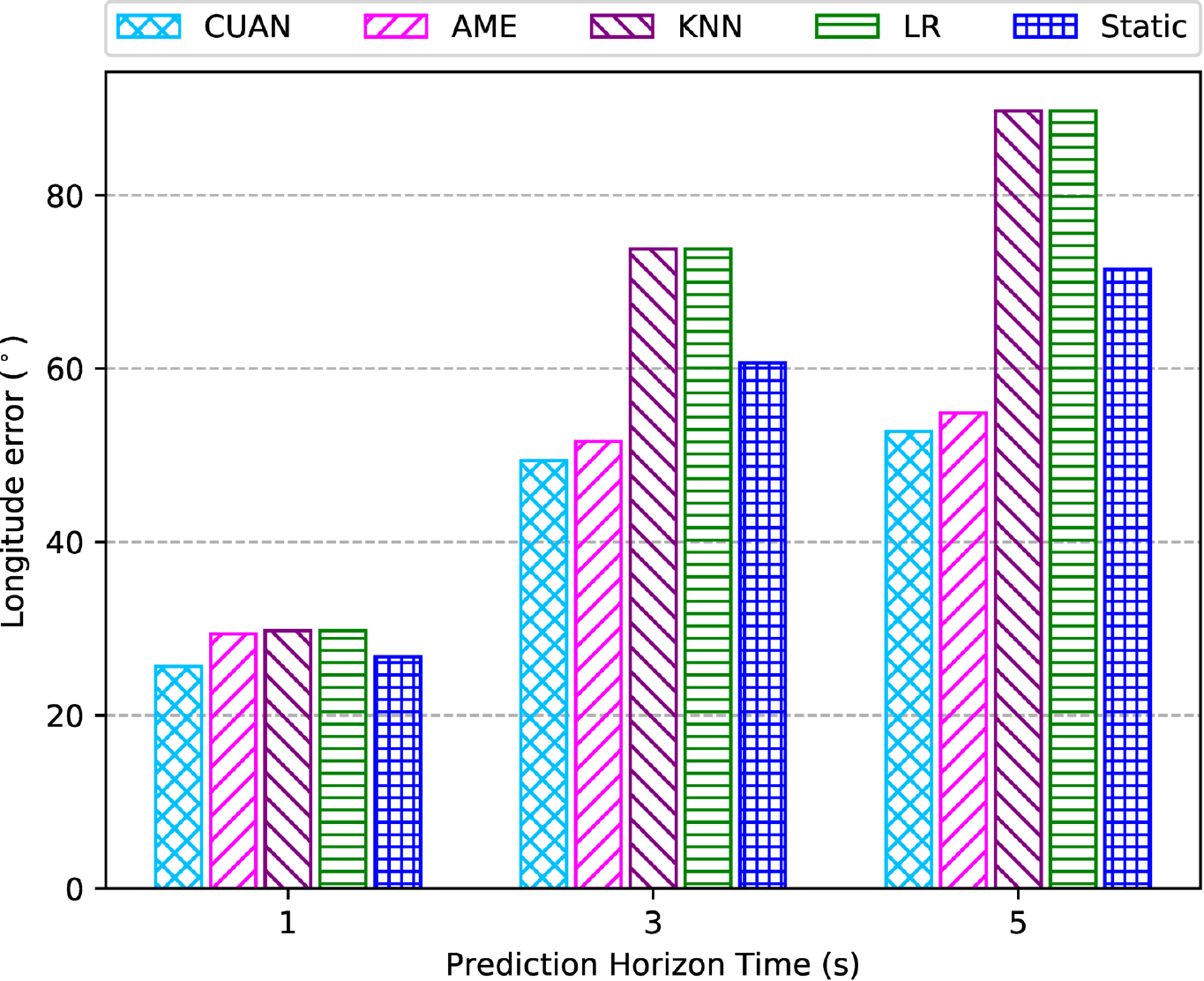}
	\end{minipage}%
	}%
    \subfigure{
	\begin{minipage}[t]{0.33\linewidth}
		\centering
		\includegraphics[width=1\linewidth]{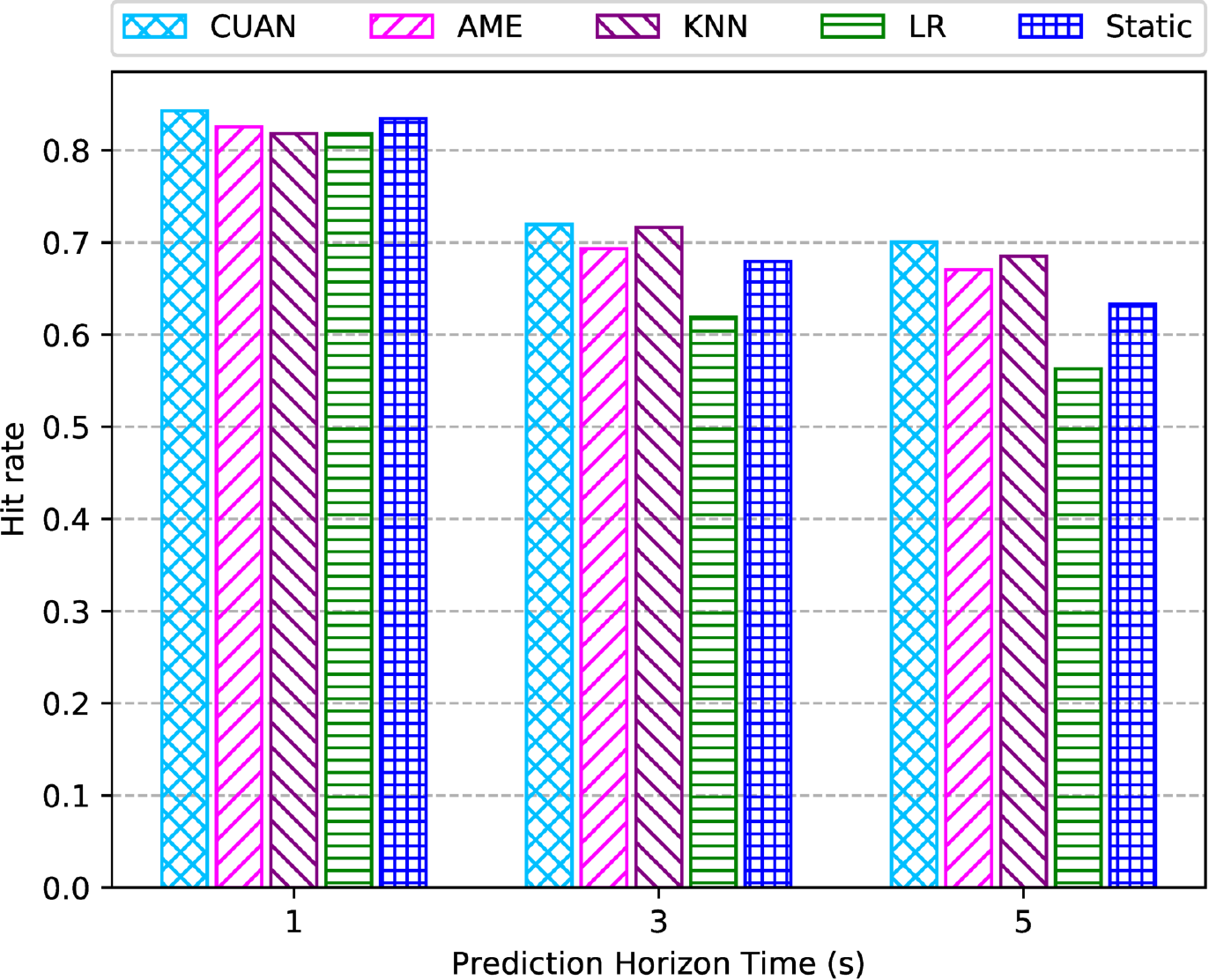}
	\end{minipage}%
	}%
    \caption{\textcolor{black}{Comparing CUAN with existing viewpoint prediction algorithms in terms of Latitude error, Longitude error, and Hit rate at the first, third, and fifth second.}}
    \label{metric}
\end{figure*}
\begin{figure*}
	\centering
	\begin{minipage}[b]{1\linewidth}
		\subfigure[]{
			\includegraphics[width=.5\linewidth]{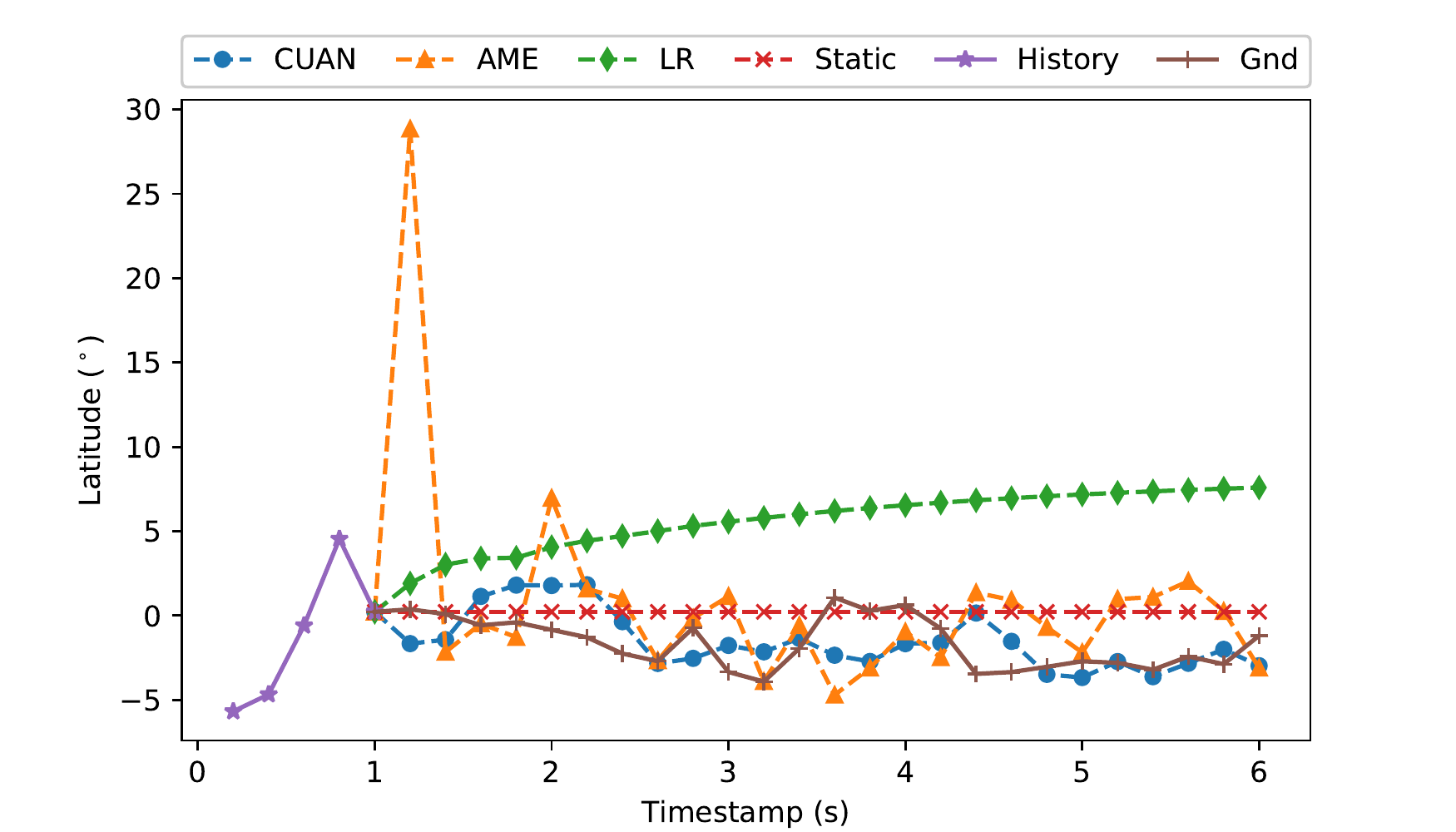} 
			\includegraphics[width=.5\linewidth]{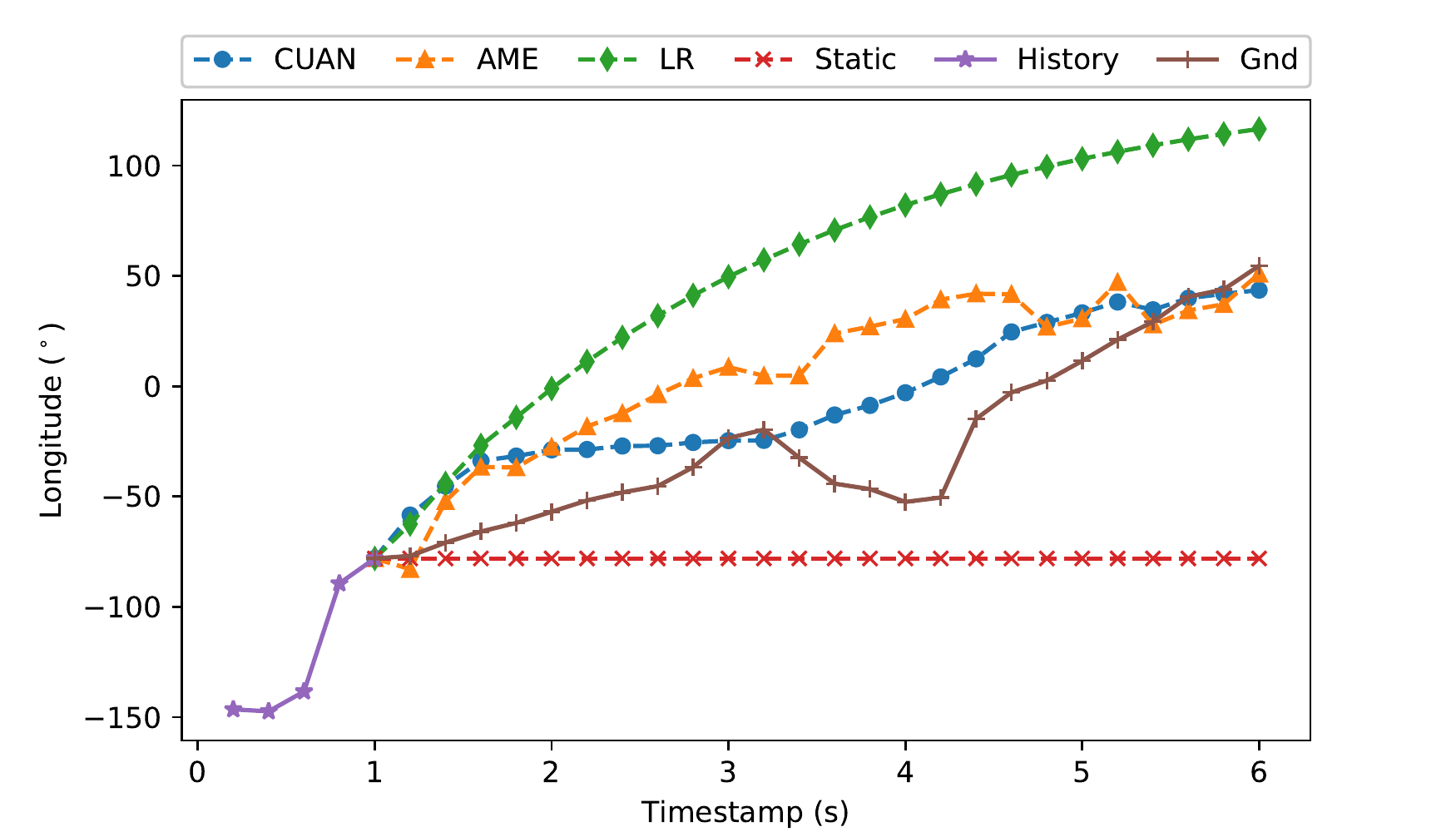} 
			\label{fig:hor_4figs_1cap_2subcap_1}
	}
	\end{minipage}
	\begin{minipage}{1\linewidth}
		\subfigure[]{
		\includegraphics[width=.5\linewidth]{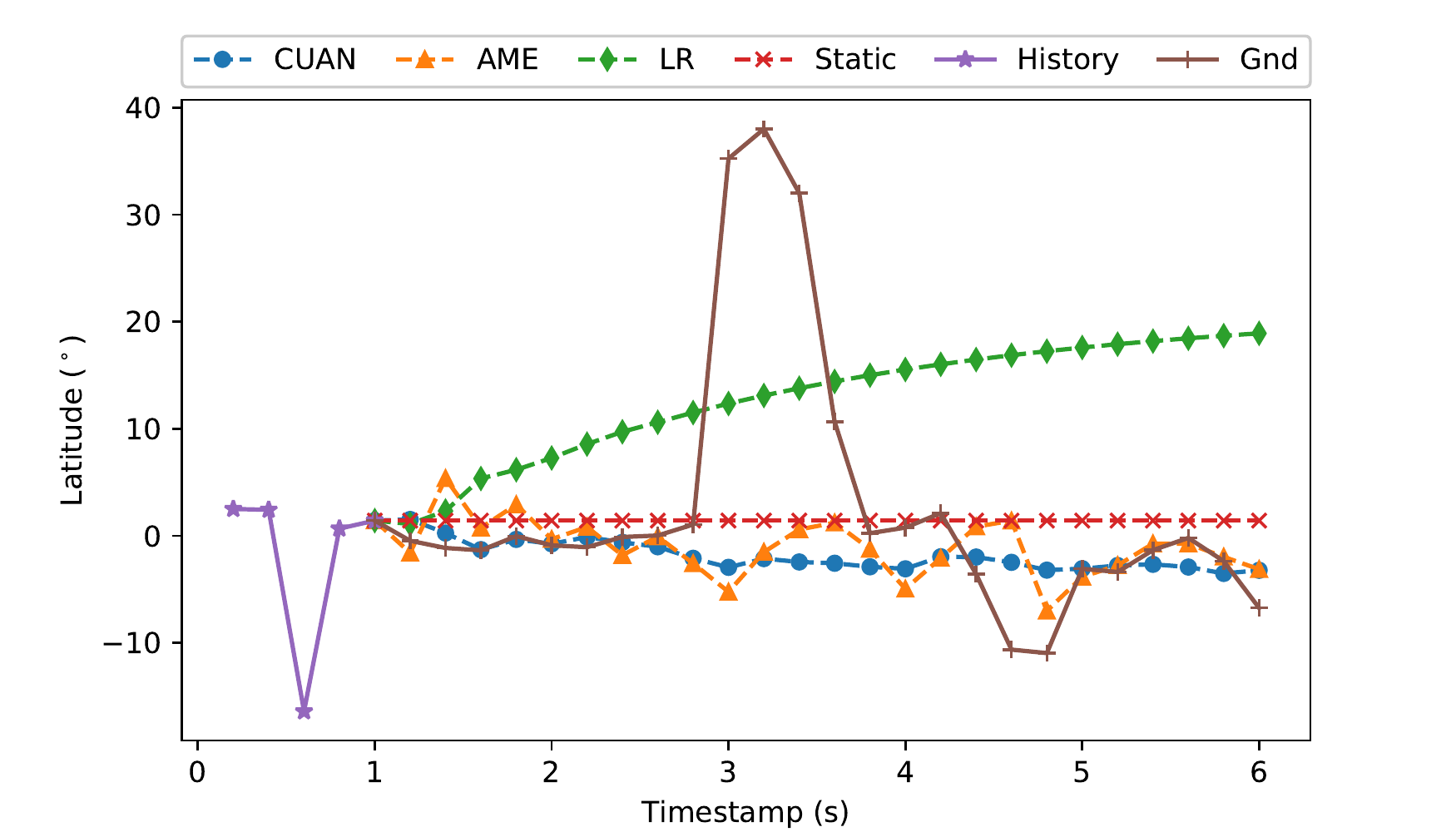} 
		\includegraphics[width=.5\linewidth]{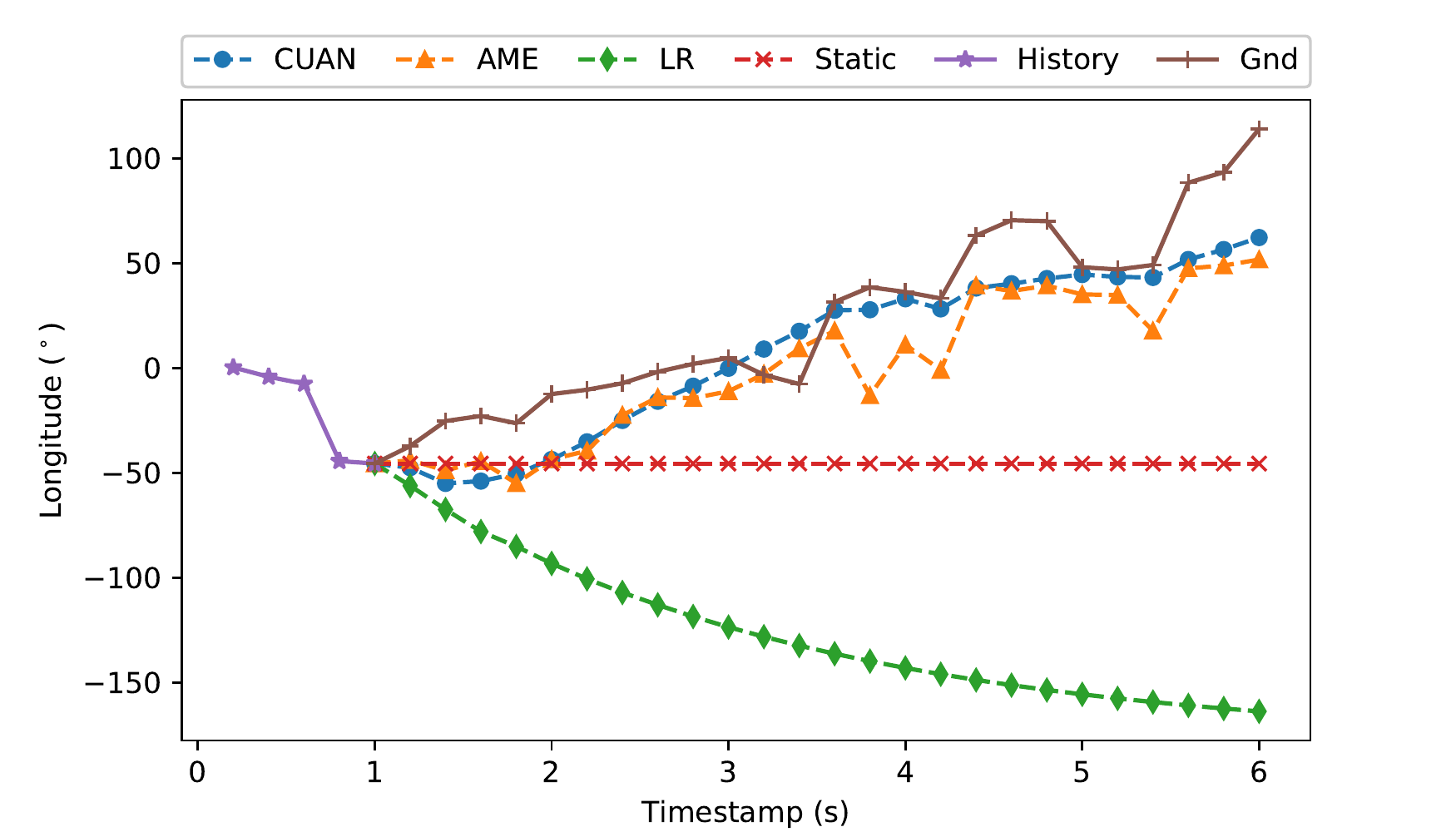} 
		\label{fig:hor_4figs_1cap_2subcap_2}
	}
	\end{minipage}
	\caption{\textcolor{black}{The visualized results of 5-second viewpoint prediction. (a) Example 1, (b) Example 2.}}
	\label{twot}
\end{figure*}
\begin{figure*}[htb]
	    \subfigure{
		\begin{minipage}[t]{0.33\linewidth}
			\centering
			\includegraphics[width=1\linewidth]{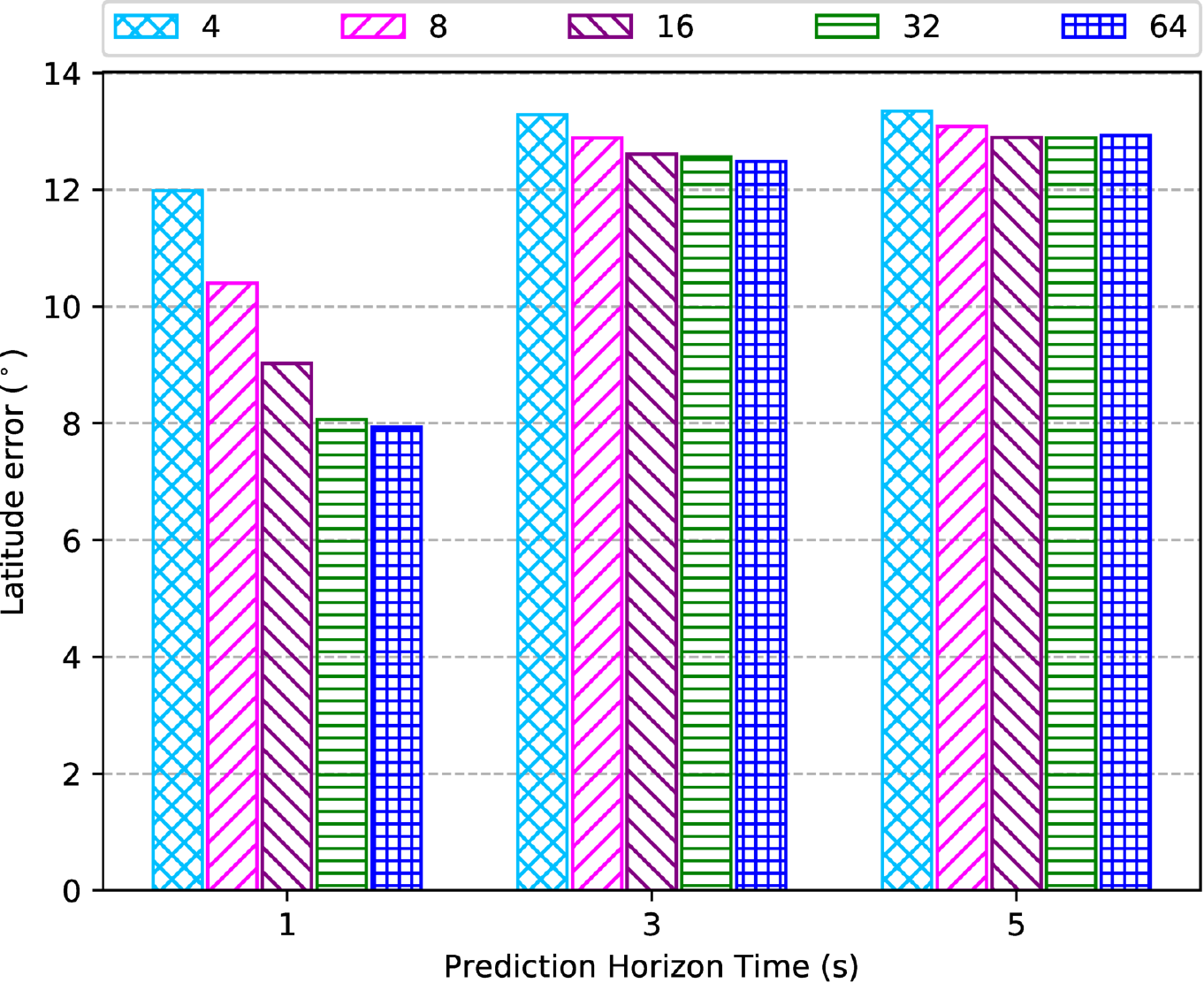}
		\end{minipage}%
	}%
	    \subfigure{
	\begin{minipage}[t]{0.33\linewidth}
		\centering
		\includegraphics[width=1\linewidth]{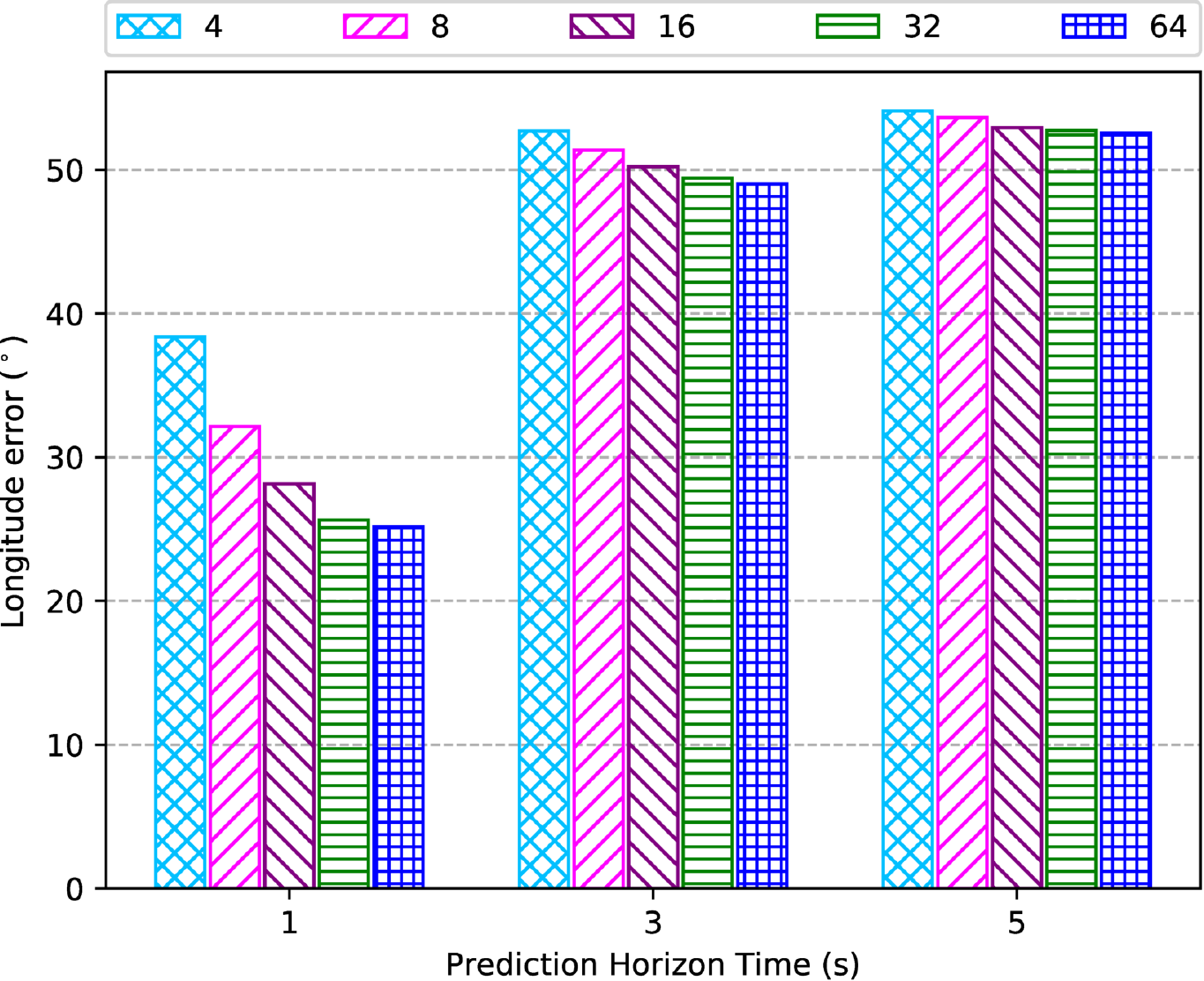}
	\end{minipage}%
	}%
	    \subfigure{
	\begin{minipage}[t]{0.33\linewidth}
		\centering
		\includegraphics[width=1\linewidth]{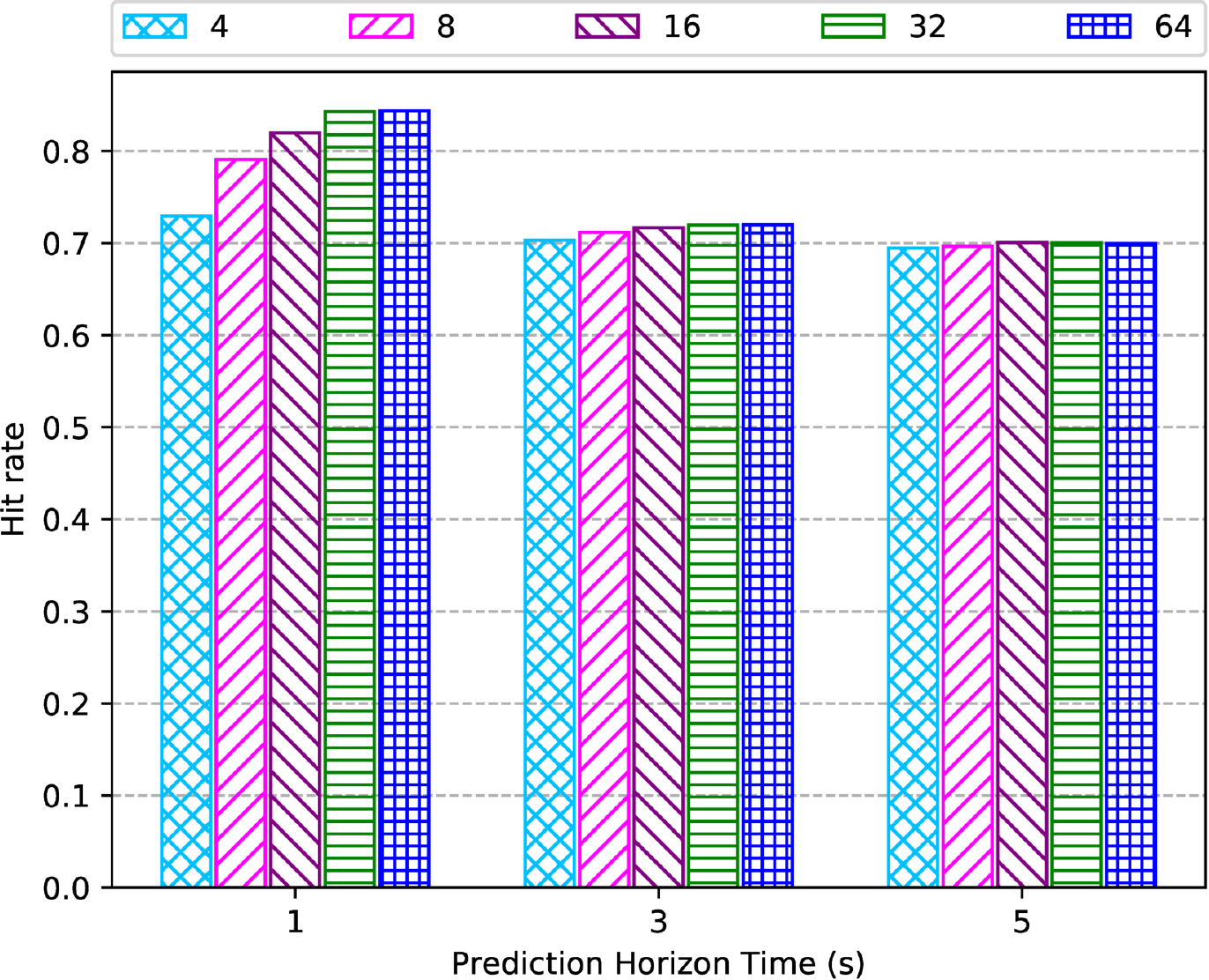}
	\end{minipage}
	}%
  \caption{\textcolor{black}{Studying the influence of the number of hidden size in terms of Latitude error, Longitude error, and Hit rate at the first, third, and fifth second.}}
  \label{runtime}
\end{figure*}
\begin{figure*}[t]
    \subfigure{
        \begin{minipage}[t]{0.33\linewidth}
        \centering
        \includegraphics[width=1.0\linewidth]{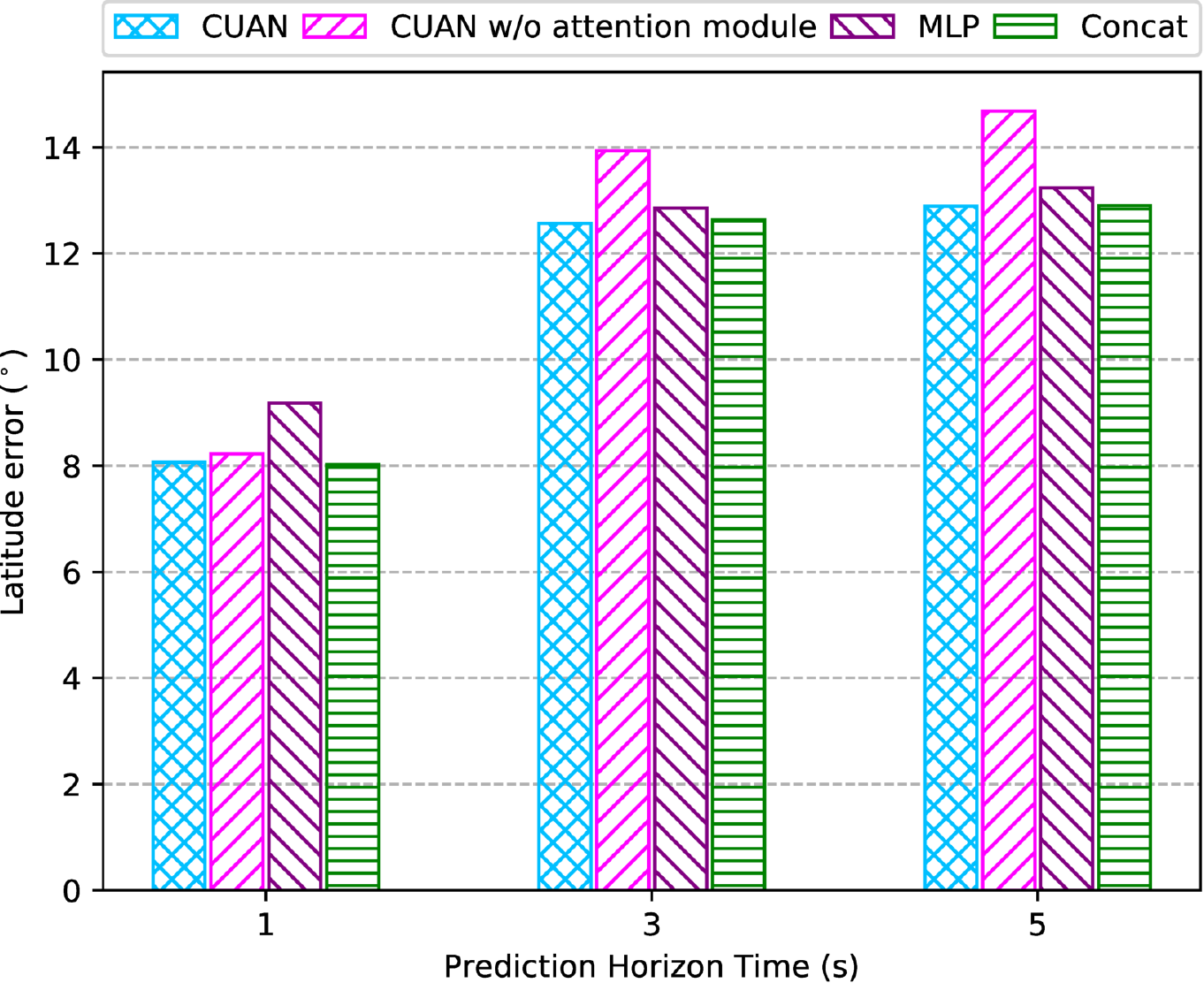}
    \end{minipage}%
    }%
    \subfigure{
        \begin{minipage}[t]{0.33\linewidth}
        \centering
        \includegraphics[width=1.0\linewidth]{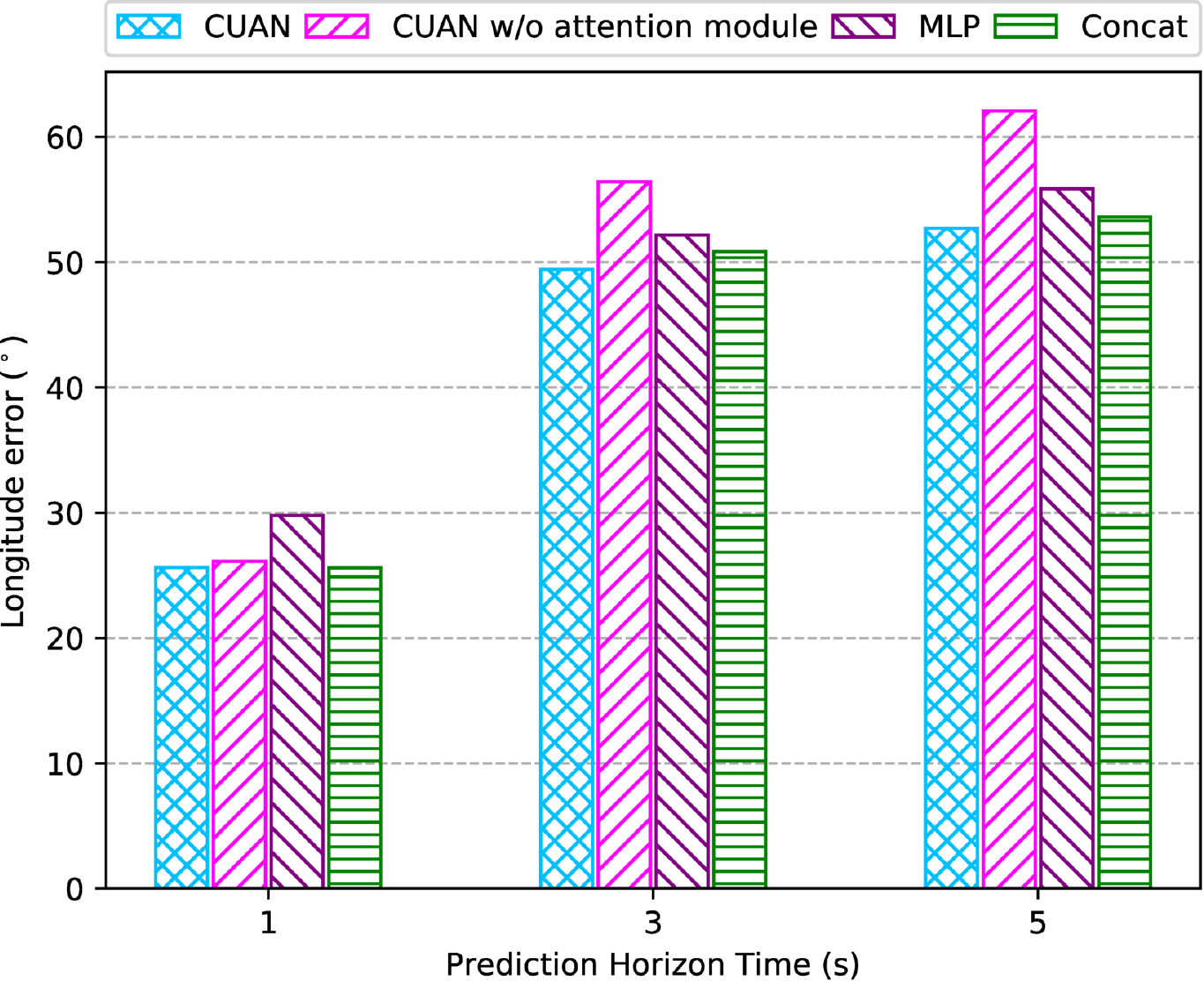}
    \end{minipage}%
    }%
    \subfigure{
        \begin{minipage}[t]{0.33\linewidth}
        \centering
        \includegraphics[width=1.0\linewidth]{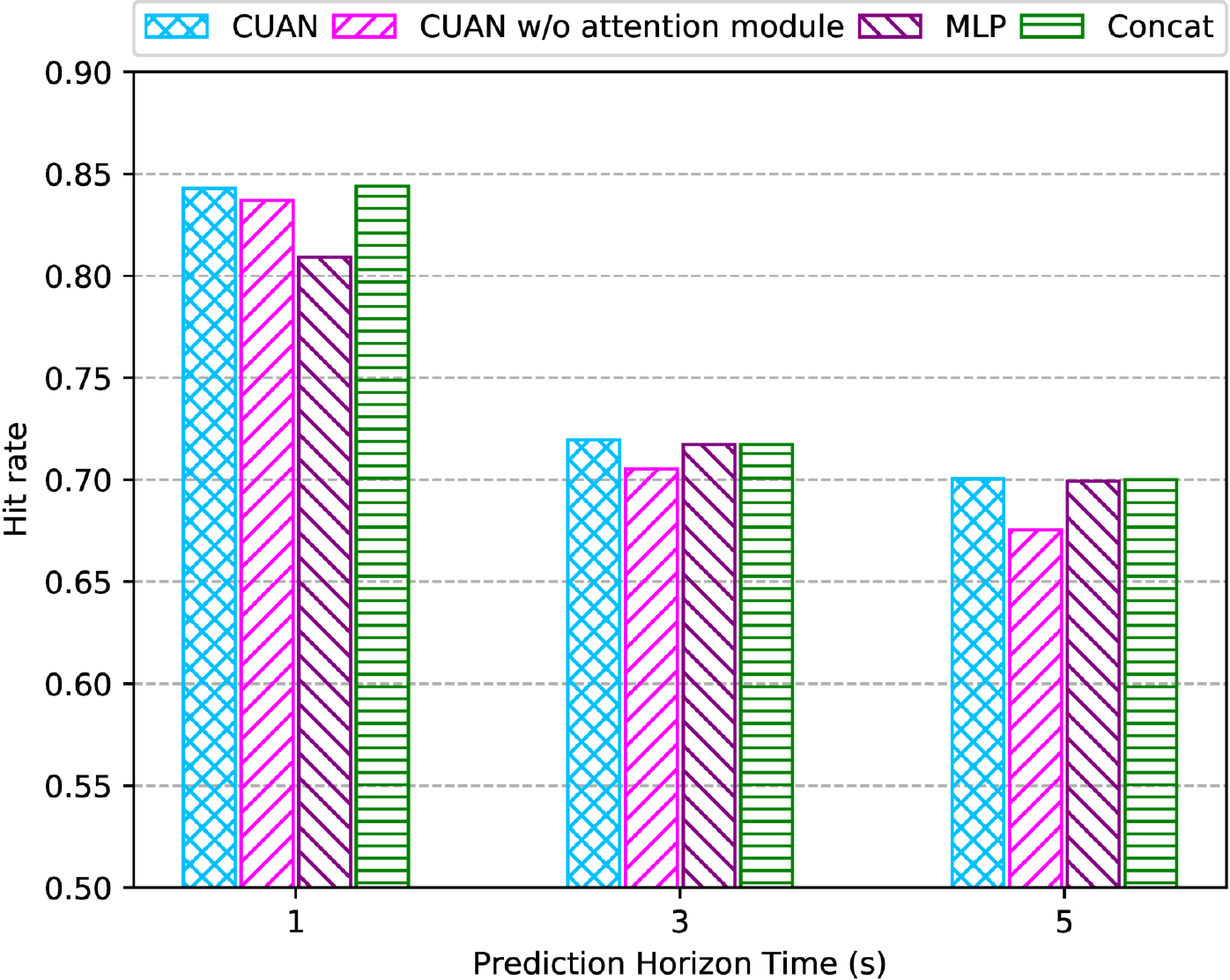}
    \end{minipage}%
    }%
    \caption{\textcolor{black}{Investigating the impact of cross-user information, self-attention mechanism, and the approach of fusing users' embeddings in terms of Latitude error, Longitude error, and Hit rate at the first, third, and fifth second. CUAN w/o attention module denotes viewpoint prediction  merely based on user's historical information, MLP denotes applying self-attention mechanism based on features extracted by multi-layer perceptron, and Concat means merging users' embeddings via concatenatation instead of addition in attention module. }}
    \label{csr}
\end{figure*}
\subsubsection{Setup}
We conduct viewpoint prediction on a widely used public dataset, i.e., the Shanghai dataset \cite{xu2018gaze}. The Shanghai dataset consists of 208 360-degree videos, where the duration of these 360-degree videos varies from 20 to 60 seconds and each video has at most 45 viewing trajectories. In this paper, we firstly normalize the longitude and latitude to be zero-centered and have range [-1, 1]. Then, we split the Shanghai dataset into training and testing sets in the same manner as \cite{xu2018gaze}. Besides, following the common setting in viewpoint prediction\cite{xu2018gaze}, we sample one frame from every five frames for model training and performance evaluation. In addition, we randomly choose 20 viewing paths as cross-user information for each video. 
 
\subsubsection{Evaluation Metrics} To evaluate the performance of the proposed method, we take the following metrics into consideration:
\begin{itemize}
    \item Longitude/Latitude Error: Longitude/Latitude Error measures the manhattan distance between the predicted longitude/latitude and its ground truth. Typically, the smaller the longitude/latitude error, the better the viewpoint prediction ability.
    \item Hit rate: In the tile-based 360-degree video streaming, it is necessary to correctly predict the viewing probabilities of tiles. As a result, Xie et al. \cite{xie2018cls} proposes Hit rate to measure the prediction precision of viewing probabilities of tiles. Typically, the higher hit rate corresponds to smaller Longitude and Latitude Error. The specific calculation of Hit rate can be found in \cite{xie2018cls}. 
\end{itemize}

\subsubsection{Methods for Comparison}
We compare our method with the following competitive baselines:
\begin{itemize}
    \item Static: This method adopts a naive strategy to forecast user' future viewpoints, i.e., assuming user' viewpoint remains still all the time.  
	\item LR \cite{xie2017360probdash}: This approach firstly estimates the trend of head movement from  user's historical viewpoints by Least Square Method, and then employs a linear regression model to predict user' future viewpoints.
	\item KNN \cite{ban2018cub360}: This scheme use the same strategy as LR to predict user' future viewpoints. To amend the prediction bias of the LR model, this method utilizes five nearest viewpoints of other users around the predicted result to calculate the viewing probabilities of tiles. 
	\item AME \cite{xie2018cls}: This approach estimates user' future viewpoints via a sequence-to-sequence learning-based model. To reduce the long-term prediction error, this method proposes an attention mechanism to exploit cross-user information. It is worth noting that AME directly fuses the predicted viewpoints with others' viewpoints, and the contribution of others' viewpoints merely depends on the similarity between user's and others' single-timestamp viewpoint.
\end{itemize}
\subsubsection{Main Results} Fig. \ref{metric} summarizes the performance of each scheme in terms of Latitude error, Longitude error,  and Hit rate in various prediction horizon time. As seen, \textcolor{black}{in terms of Hit rate at the fifth second, CUAN outperforms AME, KNN, LR, and Static by 3\%, 2\%, 14\%, and 17\%}, and shows its superior capability in decreasing longitude error and latitude error compared \textcolor{black}{to} existing methods. The similar phenomenon \textcolor{black}{can be also observed at the first and third second.}  As a result, we can conclude that, CUAN is able to better exploit cross-user information than AME, and boost the performance of KNN. In addition, we present two predicted \textcolor{black}{5-second} trajectories of each approach in Fig. \ref{twot}. As we can see, CUAN adapts better to user's head movement compared to AME, while LR is easily biased since the assumption of linear head movement is easily violated. 
\begin{table}[t]
	\caption{\textcolor{black}{the average runtime of cuan at different length of viewpoint prediction.}}
	\centering
	\begin{tabular}{|c|c|c|}
		\hline
		\multicolumn{1}{|l|}{\multirow{2}{*}{Length of viewpoint prediction (s)}} & \multicolumn{2}{c|}{Average Runtime (ms)}                \\ \cline{2-3} 
		\multicolumn{1}{|l|}{}                                             & \multicolumn{1}{c|}{CPU} & \multicolumn{1}{c|}{GPU} \\ \hline
        1      &  8.9  &   7.7 \\ \hline
		3      &  20.1 & 19.7 \\ \hline
		5      &  32.9 & 32.4  \\
		\hline
	\end{tabular}
	\label{CUAN:runtime}		
\end{table}
\subsubsection{Ablation Study} 
Firstly, we vary the hidden size of LSTMs from 4 to 64. Results from this sweep are presented in Fig. \ref{runtime}. As shown, the performance of CUAN is increasingly improved with the increase of hidden size, and begins to plateau once the hidden size exceeds 32. Hence, we set the hidden size of LSTMs as 32, \textcolor{black}{and keep the same configuration in the following experiments}. \textcolor{black}{Then, we investigate the impact of cross-user information, self-attention mechanism (LSTM-based or MLP-based), and the approach of fusing users' embeddings (addition or concatenation) on the performance of viewpoint prediction. As seen in Fig.8, compared to CUAN without cross-user information, methods equipped with cross-user information show their superior capability in forecasting users' future viewpoints (especially in the long-term prediction). Meantime, we can observe that MLP-based self-attention is inferior to LSTM-based CUAN in terms of Latitude error, Longitude error, and Hit rate, which is mainly because LSTM is more suitable to capture sequential information than MLP. Also, we can find adding users' embeddings performs slightly better than concatenating users' embeddings in the task of viewpoint prediction. Therefore, we choose to merge users' embeddings through addition instead of concatenation in CUAN. Finally, we evaluate the average runtime of CUAN at different length of viewpoint prediction. As shown in Table \ref{CUAN:runtime}, CUAN meets the real-time requirements of practical applications.}

\subsection{Rate Adaptation}
\subsubsection{Setup}
We evaluate the performance of the proposed 360SRL in the following setting: 
\begin{itemize}
    \item Video Content:  We randomly select 30 and 18 360-degree videos from the Shanghai dataset \cite{xu2018gaze} as our training and testing data. To achieve adaptive streaming, we split each 360-degree video in the $3 \times 3$ (rows$\times$columns) tiling pattern, and encode each tile with a quantization parameter (QP) of 22 to 42 in steps of 5 via the open-source encoder x264 \cite{merritt2006x264}, and then encapsulate the encoded bit-streams into chunks recording 1-second video content.
    
    \item Network Traces: We randomly select 50 traces from the HSDPA dataset \cite{riiser2013commute} as training data, and the remaining 28 traces as testing. Noticing that the network capacity is often unaffordable to pick up the minimum bitrate, we enlarge the network capacity by 3 Mbps. Meanwhile, in order to avoid the situation where the network capacity can always afford to pick up the maximum bitrate, we set the upper bound of the network capacity to 8 Mbps.

    \item Viewpoint Traces: Since the Shanghai dataset \cite{xu2018gaze} has collected multiple viewing trajectories for each video, we directly reuse those data.

\end{itemize}
 All experiment is conducted on a NVIDIA TITAN X GPU platform with an Intel(R) Core(TM) i7-6850K 3.60GHz CPU.

\begin{figure*}[htbp]
    \subfigure[]{
        \begin{minipage}[t]{0.33\linewidth}
        \centering
        \includegraphics[width=1\linewidth]{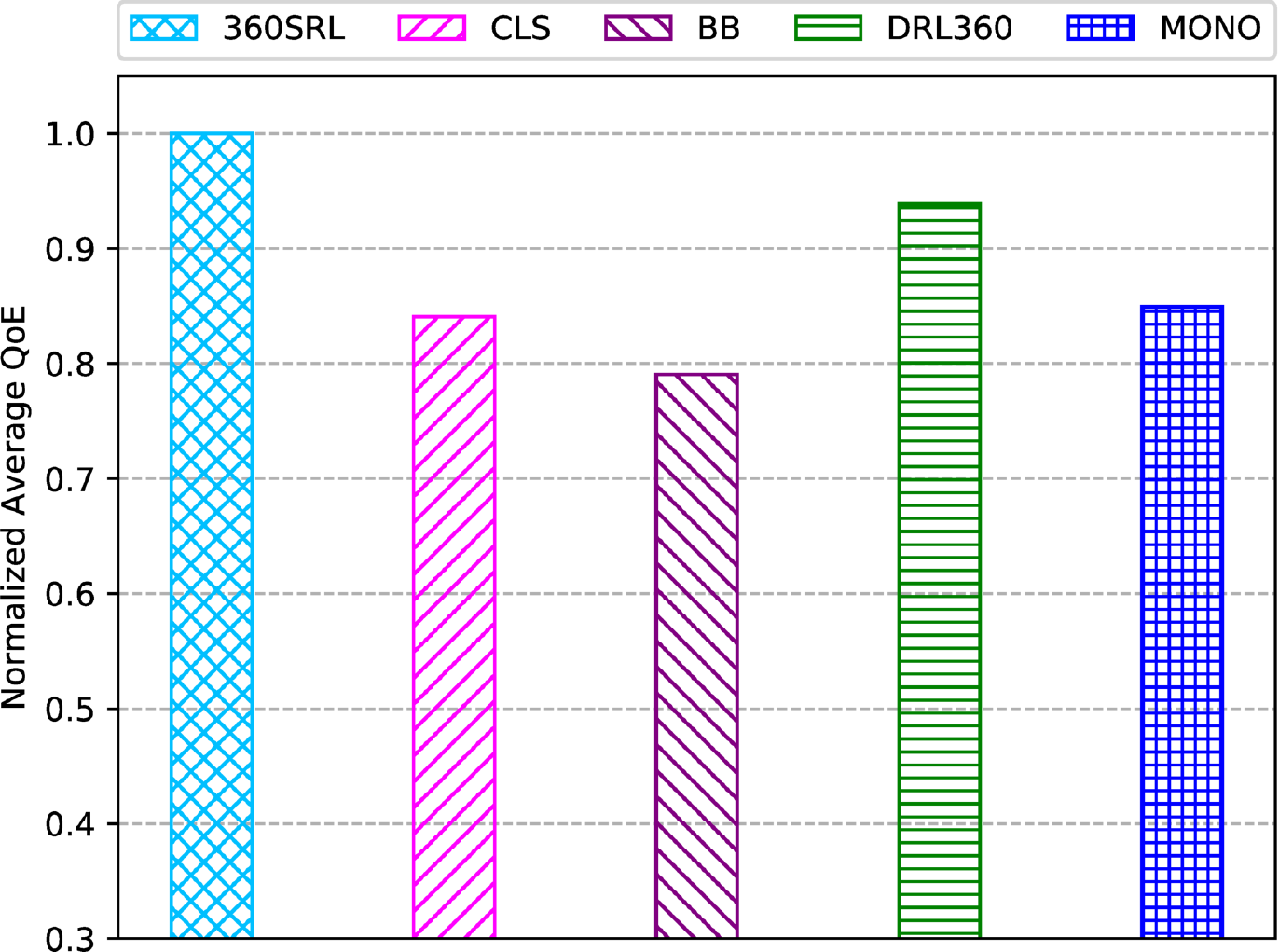}
    \end{minipage}%
    }%
    \subfigure[]{
        \begin{minipage}[t]{0.33\linewidth}
        \centering
        \includegraphics[width=1\linewidth]{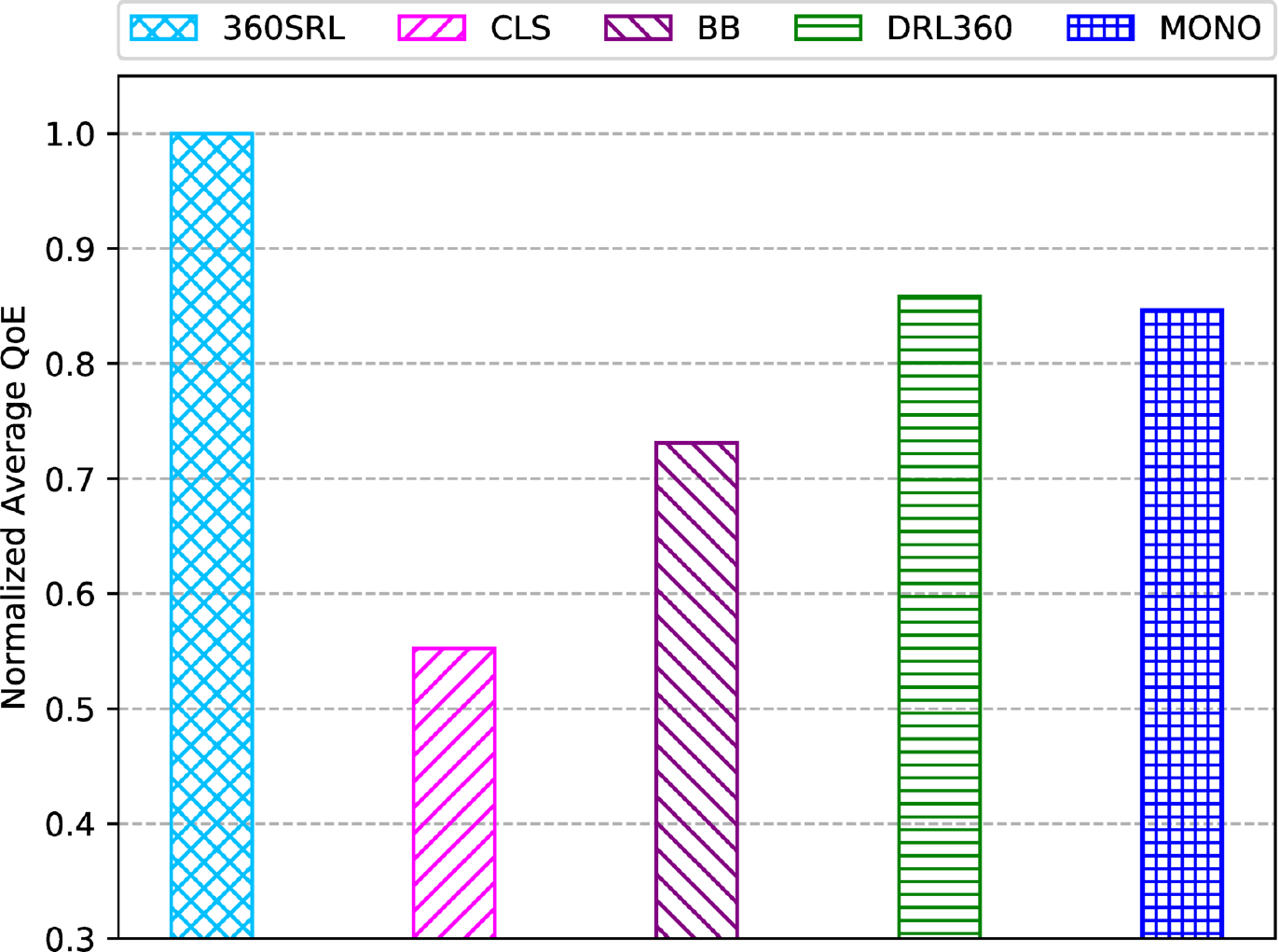}
    \end{minipage}%
    }%
    \subfigure[]{
        \begin{minipage}[t]{0.33\linewidth}
        \centering
        \includegraphics[width=1\linewidth]{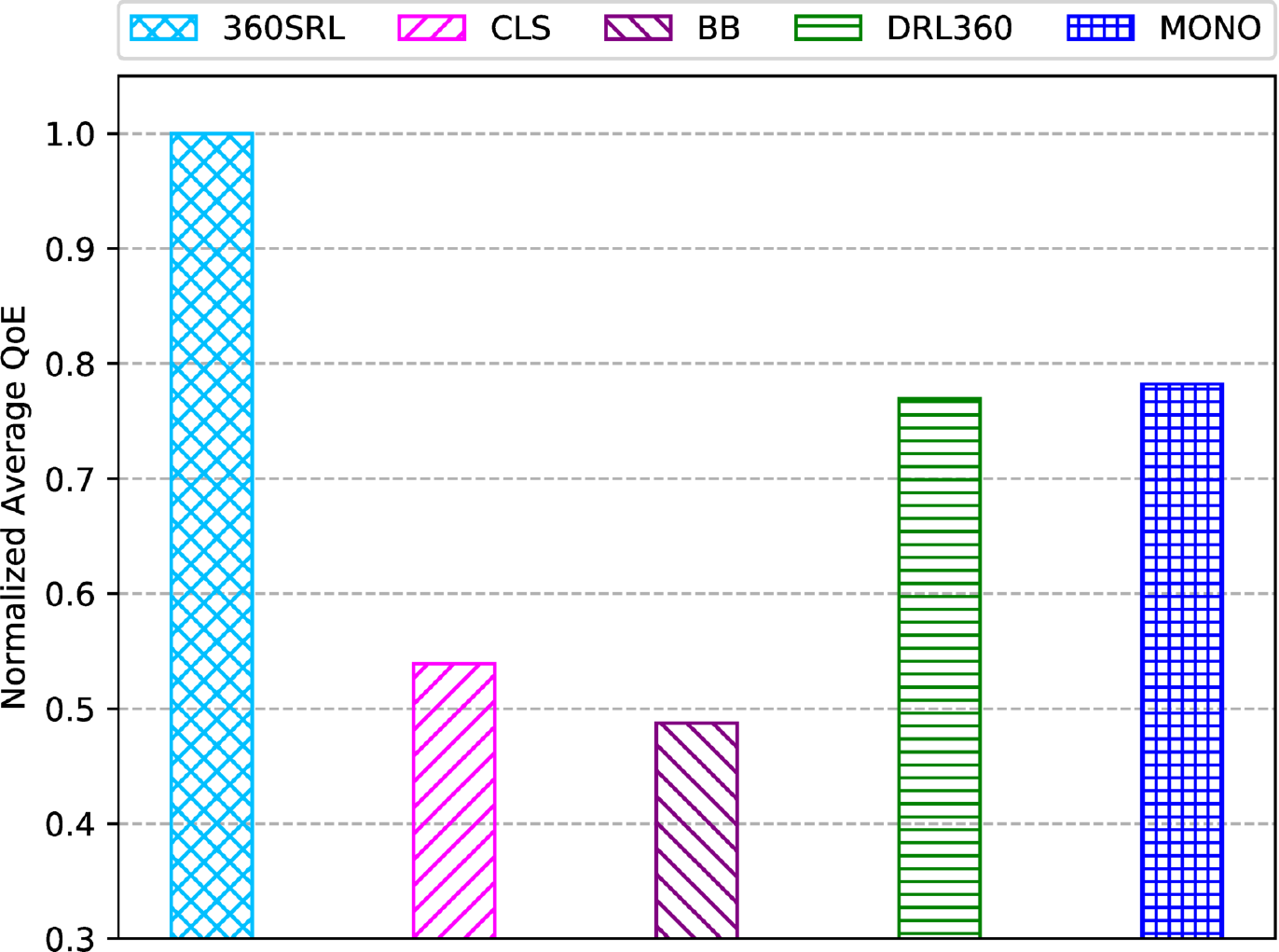}
    \end{minipage}%
    }%
    \caption{\textcolor{black}{Comparing 360SRL with existing ABR algorithms on three different QoE objectives: a) $\eta_1=1,\eta_2=1,\eta_3=4.3$, 2) $\eta_1=1,\eta_2=2,\eta_3=4.3$, c) $\eta_1=2,\eta_2=1,\eta_3=4.3$.}}
    \label{average qoe}
\end{figure*}

\begin{figure*}[htbp]
    \subfigure{
        \begin{minipage}[t]{0.25\textwidth}
        \centering
        \includegraphics[width=1\textwidth]{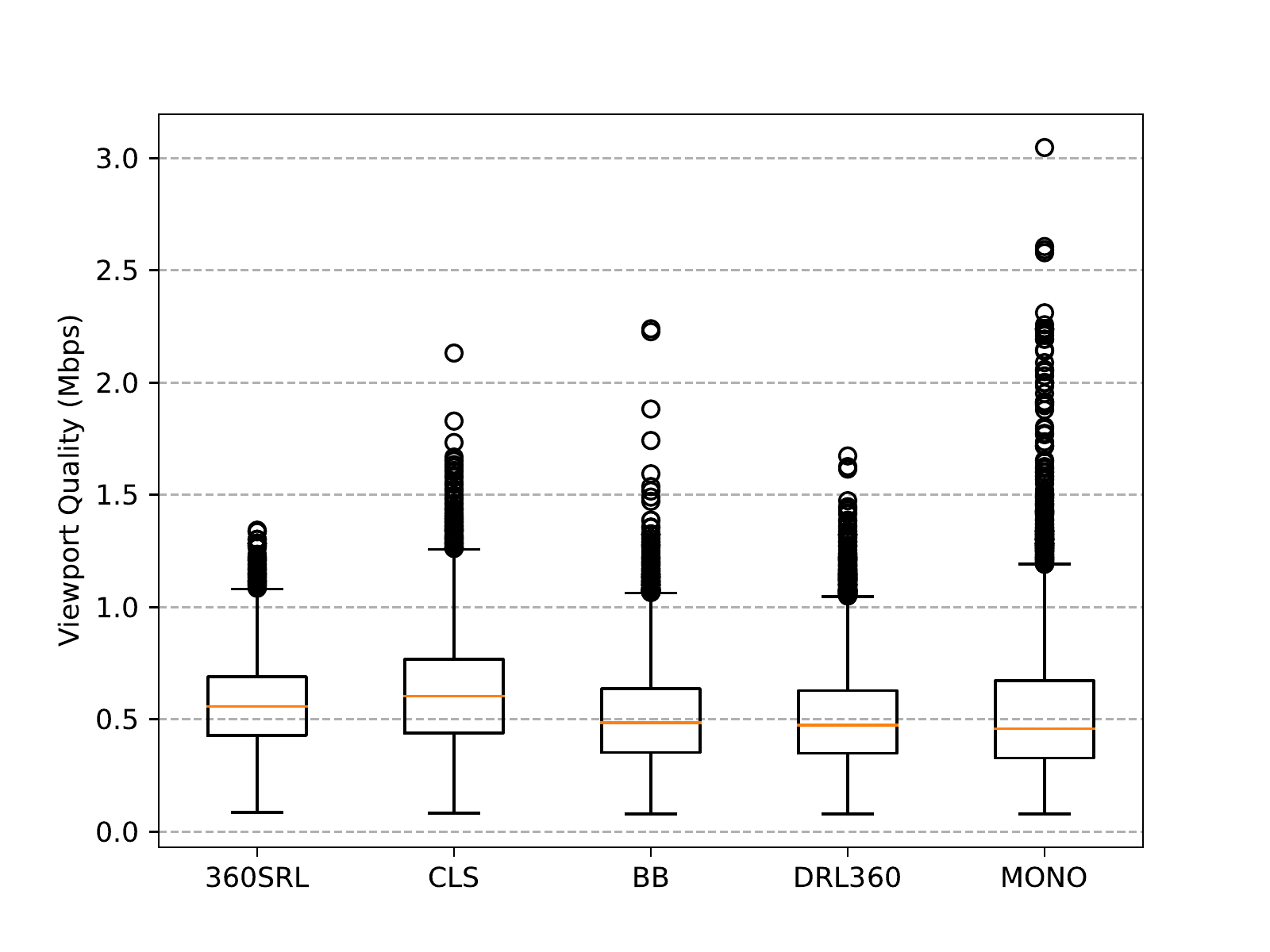}
    \end{minipage}%
    }%
    \subfigure{
        \begin{minipage}[t]{0.25\linewidth}
        \centering
        \includegraphics[width=1\linewidth]{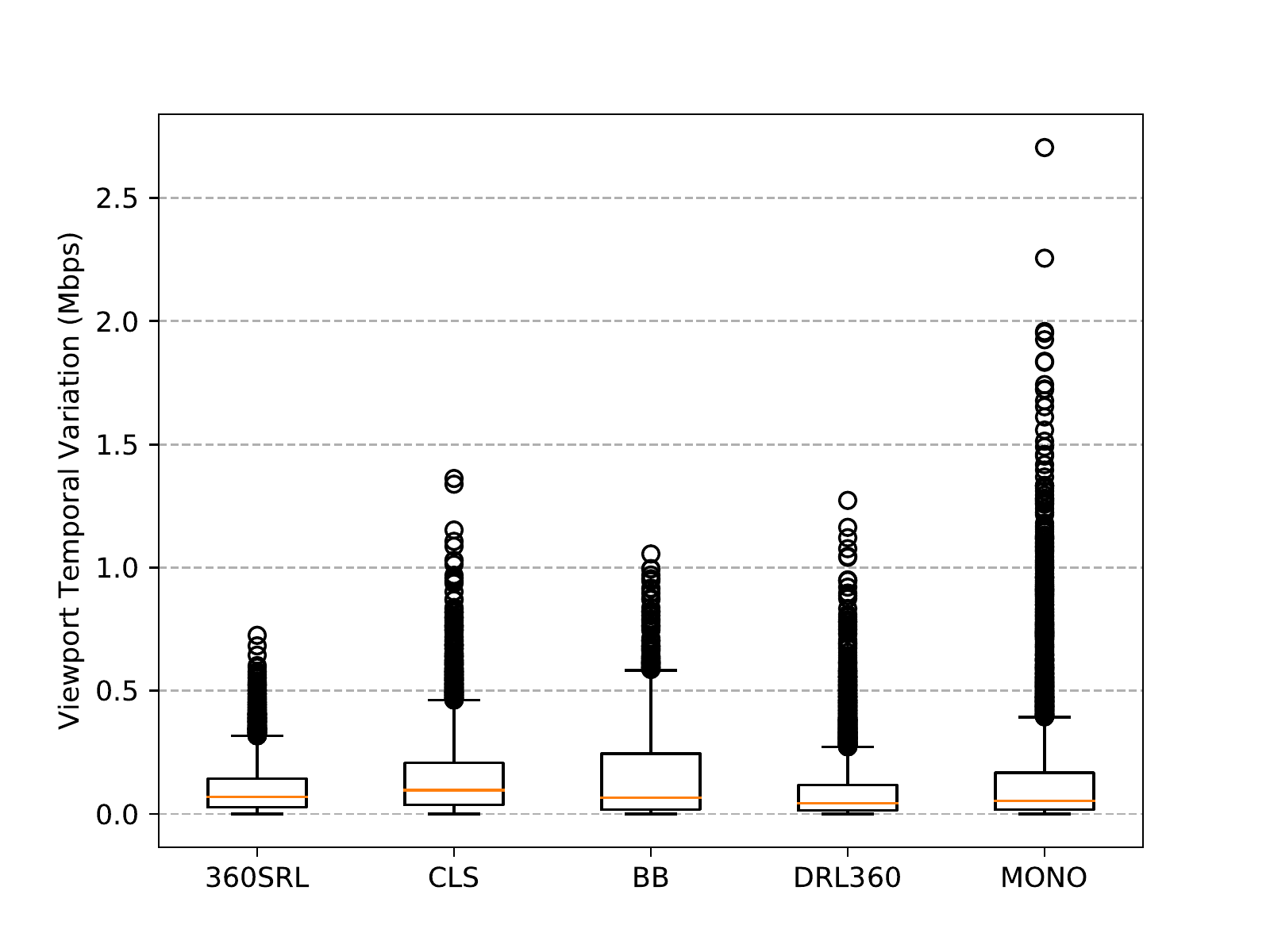}
    \end{minipage}%
    }%
    \subfigure{
        \begin{minipage}[t]{0.25\linewidth}
        \centering
        \includegraphics[width=1\linewidth]{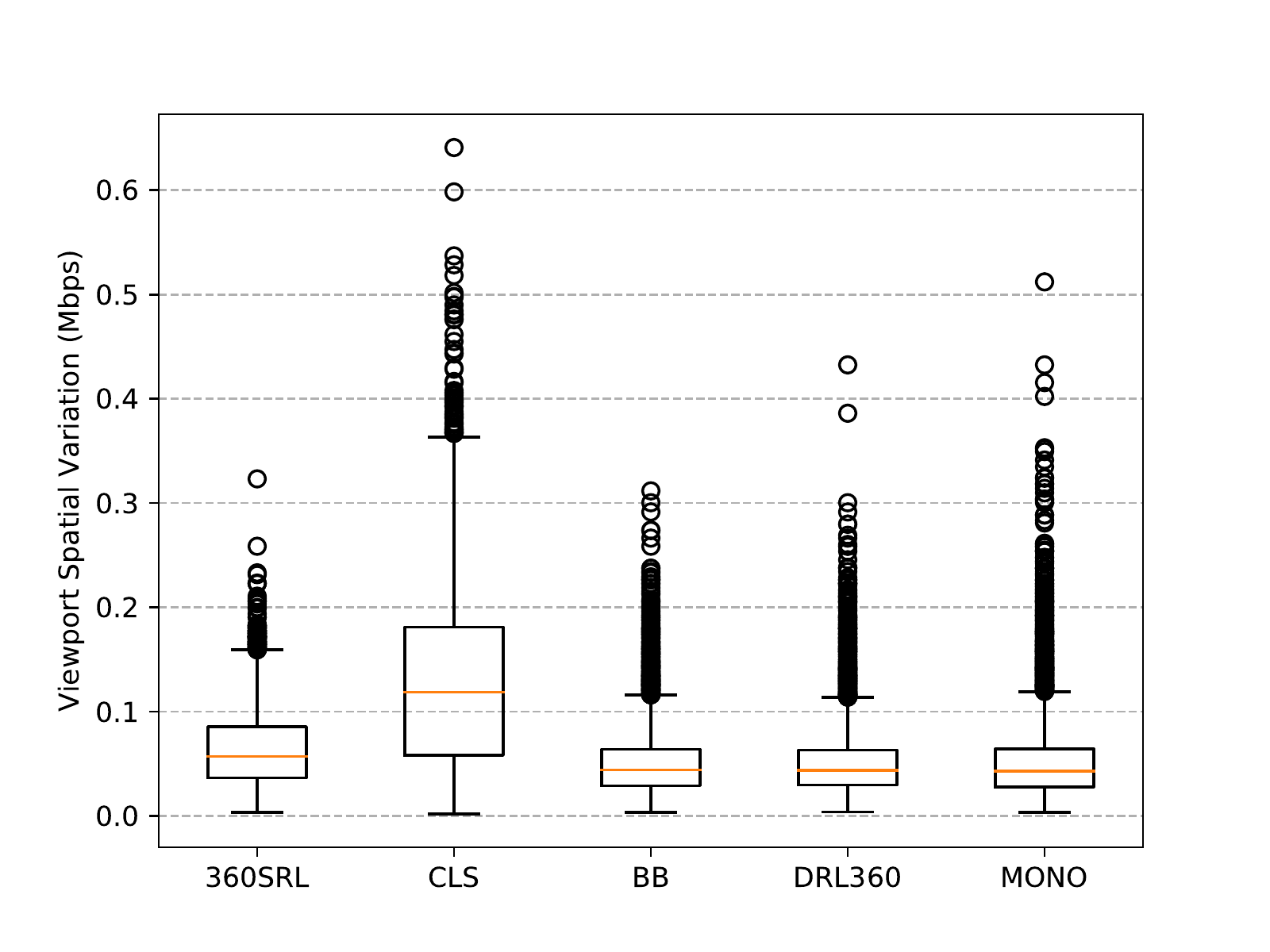}
    \end{minipage}%
    }%
    \subfigure{
        \begin{minipage}[t]{0.25\linewidth}
        \centering
        \includegraphics[width=1\linewidth]{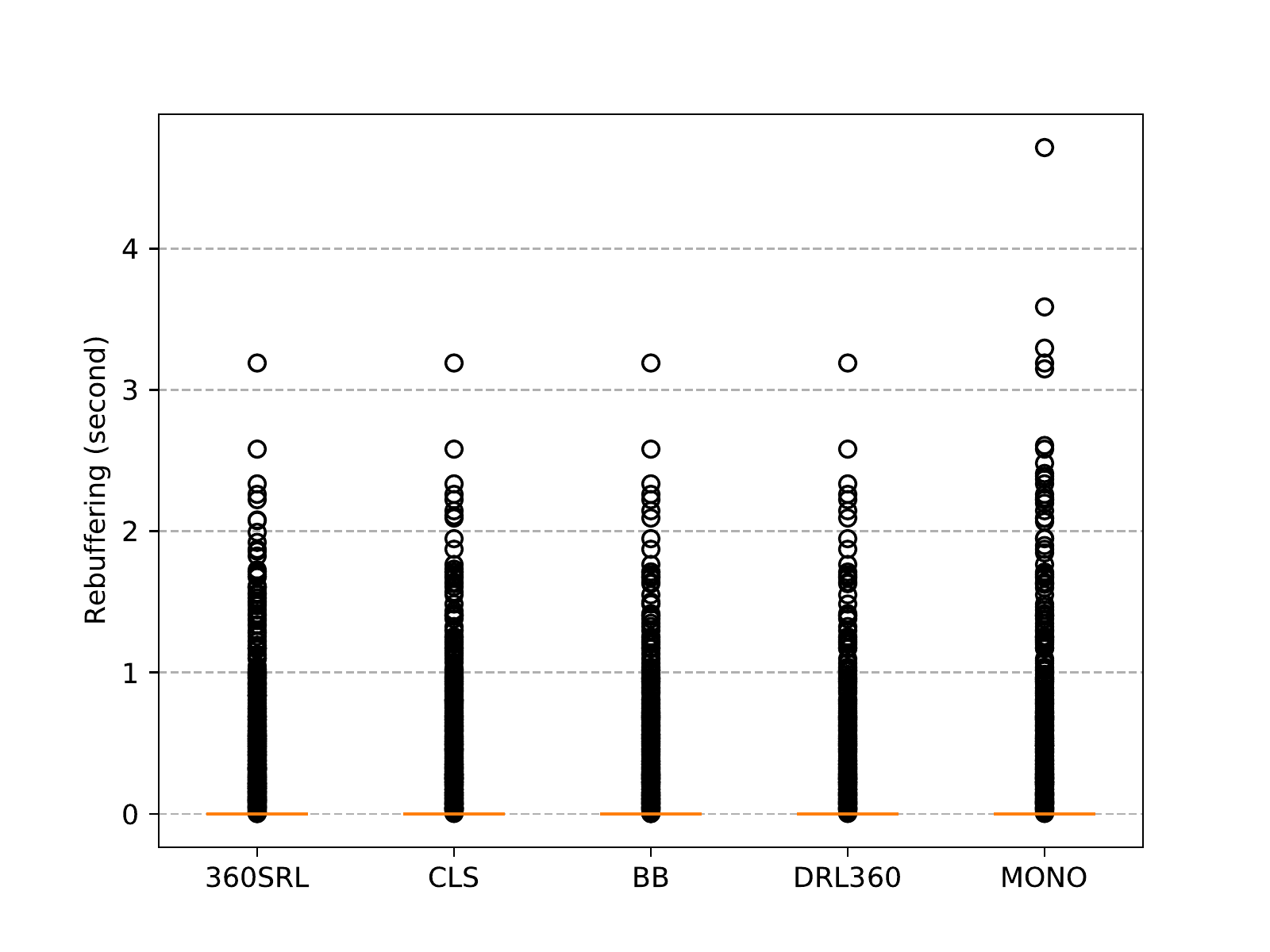}
    \end{minipage}%
    }%
  \caption{\textcolor{black}{Comparing 360SRL with existing ABR algorithms in terms of viewport quality, viewport temporal quality variance, viewport spatial quality variance, and rebuffering when the configuration of the QoE objective is $\eta_1=1,\eta_2=1,\eta_3=4.3$.}}

    \label{vpsnr}
\end{figure*}

\subsubsection{Methods for Comparison}
We compare our method with the following methods which collectively represent the state-of-the-art in rate adaptation: 
\begin{itemize}
	\item MONO: This approach regards the panoramic video streaming as ordinary video streaming, and employs the RL-based ABR algorithm proposed by Mao et al. \cite{mao2017neural} to pick up the bitrate for the upcoming segments.
	\item Buffer-based (BB): This scheme selects the same bitrate for tiles within user's FoV via the buffer-based algorithm proposed by Huang et al. \cite{huang2015buffer}, which uses a reservoir of 1 seconds and a cushion of 5 seconds i.e., it chooses bitrates with the goal of maintaining the buffer occupancy above 1 seconds, and automatically picks up the highest available bitrate if the buffer occupancy is larger than 5 seconds. It is worth noting that tiles out of user's FoV are delivered at the lowest bitrate. 
	\item CLS \cite{xie2018cls}: This method firstly utilizes the target-buffer based algorithm predict the bandwidth budget, and then conducts rate allocation according to the viewing probabilities of tiles. It is worth noting that this method aims to maximize the viewpoint quality regardless of the viewpoint temporal and spatial variation. 
	\item DRL360 \cite{zhang2019drl360}: This approach employs a RL-based algorithm to select the same bitrate for tiles within user's FoV, and picks up the lowest bitrate for the remaining tiles. 
\end{itemize}
To keep fair comparison, all ABR schemes use the proposed CUAN to predict user's future viewpoints. 

\subsubsection{Main Results}
As illustrated in Fig.\ref{average qoe}, 360SRL achieves the state-of-the-art performance under different QoE objectives, which validates the advantage of 360SRL. Specifically, in the setting of $\eta_1=1,\eta_2=1,\eta_3=4.3$, 360SRL outperforms MONO, CLS, BB, and DRL360 by 15\%, 16\%, 21\%, and 6\% in terms of normalized average QoE, respectively. When users prefer to small veiwport temporal variation and viewport spatial variation, 360SRL is also superior to the existing algorithms. To figure out the improvement in the average QoE obtained by 360SRL, we analyze the performance of each scheme on the individual terms of the definition of QoE score (Eq. \ref{hello}). As seen in Fig. \ref{vpsnr}, in the case of $\eta_1=1,\eta_2=1,\eta_3=4.3$, 360SRL is comparable to CLS in terms of viewpoint quality and rebuffering, while shows superior capability in suppressing viewpoint temporal and spatial variation. The similar phenomena also occurs on the other two QoE objectives. Hence, we can conclude that the advanced capability in reconciling each individual item of Eq. \ref{hello} helps 360SRL beat the existing competitive ABR algorithms. In addition, compared with DRL360, we can observe that a large portion of the gain obtained by 360SRL comes from its higher viewport quality. This is mainly because 360SRL automatically chooses bitrate for tiles within user's FoV according to their viewing probabilities, instead of simply selecting the same bitrate for them.

\subsubsection{Ablation Study} \textcolor{black}{Firstly, to investigate the impact of parametric size of 360SRL on the average QoE, we vary the number of filters in the 1D-CNN and the number of neurons in the FC from 4 to 128. As shown in Table \ref{drl:hidden size}, the average QoE begins to plateau once the number of filters and neurons reaches 64. Thus, we set the number of filters and neurons as 64 and keep it unchanged in the following experiments. In addition, we can notice that the performance of 360SRL is relatively insensitive to the parametric size: the average QoE with 4 filters and neurons is within 1.9\% of that when the number of filters and neurons is 64. Secondly, we study the impact of training algorithm of 360SRL on the average QoE. As shown in Table. \ref{drl:training algorithm}, A3C achieves better performance than DQN \cite{fu2019360srl} at three different configurations of QoE objective. As a result, in this paper, we switch the implement of 360SRL from DQN to A3C without introducing extra challenges.} Thirdly, we investigate the impact of maximum buffer occupancy of the video player on the average QoE. As seen in Table. \ref{table:maximum buffer occupancy}, the maximum buffer occupancy of the video player has negligible impact on the average QoE. the average QoE with a 2-second maximum buffer occupancy is within 0.1\% of that when the maximum buffer occupancy is 10 seconds. \textcolor{black}{Fourthly}, we study the effect of decision order on the average QoE. Table. \ref{order} summaries the performance of Z-scan order, random order, the order of viewing probability from low to high, and the order of viewing probability from high to low. As seen, 360SRL is sensitive to the decision order. This may be because it is hard for 360SRL to react when a single dimension of the input is changed. Hence, we adopt a greedy strategy, i.e., the order of viewing probability from high to low, as our decision order. \textcolor{black}{Fifthly}, we probe the impact of viewpoint prediction on the average QoE. As illustrated in Table. \ref{VP}, CUAN achieves a significant improvement in average QoE, compared with existing ABR algorithms, which validates the effectiveness of CUAN. \textcolor{black}{Finally, we also evaluate the average runtime of 360SRL on CPU and GPU platform. As seen in Table \ref{drl:runtime}, 360SRL achieves the real-time requirements of real-world applications.}
\begin{table}[htbp]
	\caption{\textcolor{black}{the impact of number of filters and neurons on the average qoe when the configuration of the qoe objective is $\eta_1=1,\eta_2=1,\eta_3=4.3$.}}
	\centering
	\begin{tabular}{|c| c|}
		\hline
		Number of filters and neurons (each) &  Average QoE \\
		\hline
		4 & 12.17  \\
		\hline
		8 &   12.25 \\
		\hline
		16 &  12.16 \\
		\hline
		32 &  12.26 \\
		\hline
		64 &  12.40 \\
		\hline
		128 &  12.34 \\			
		\hline
	\end{tabular}
	\label{drl:hidden size}		
\end{table}

\begin{table}[htbp]
	\caption{\textcolor{black}{the impact of training algorithm on the average qoe.}}
	\centering
	\begin{tabular}{|c|c|c|}
		\hline
		\multicolumn{1}{|l|}{\multirow{2}{*}{QoE objective}} & \multicolumn{2}{c|}{Method}                \\ \cline{2-3} 
		\multicolumn{1}{|l|}{}                                             & \multicolumn{1}{c|}{DQN} & \multicolumn{1}{c|}{A3C}       \\ \hline
		$\eta_1=1,\eta_2=1,\eta_3=4.3$  &           11.83          &      12.40                    \\ \hline
		$\eta_1=1,\eta_2=2,\eta_3=4.3$  &           10.37          &       11.00                  \\ \hline
		$\eta_1=2,\eta_2=1,\eta_3=4.3$  &           9.71           &      10.23                   \\
		\hline
	\end{tabular}
	\label{drl:training algorithm}		
\end{table}

\begin{table}[htbp]
	\caption{\textcolor{black}{the impact of maximum buffer occupancy of the video player on the average qoe when the configuration of the qoe objective is $\eta_1=1,\eta_2=1,\eta_3=4.3$.}}
	\centering
	\begin{tabular}{|c| c|}
		\hline
		Maximum Buffer Occupancy (second) &  Average QoE \\
		\hline
		2 & 12.302  \\
		\hline
		4 & 12.395 \\
		\hline
		6 & 12.397 \\
		\hline
		8 &  12.397 \\
		\hline
		10 & 12.398 \\
		\hline
	\end{tabular}
	\label{table:maximum buffer occupancy}		
\end{table}
\begin{table}[htbp]
	\caption{\textcolor{black}{the impact of viewpoint prediction on the average qoe when the configuration of the qoe objective is $\eta_1=1,\eta_2=1,\eta_3=4.3$.}}
	\centering
	\begin{tabular}{|c| c|}
		\hline
		Method  &  Average QoE \\
		\hline
		Static & 11.30  \\
		\hline
		LR &   9.62 \\
		\hline
		KNN &  11.64 \\
		\hline
		AME &  11.79 \\
		\hline
		CUAN &  12.40 \\		
		\hline
	\end{tabular}
	\label{VP}		
\end{table}
\begin{table}[htbp]
	\caption{\textcolor{black}{the impact of decision  order on the average qoe when the configuration of the qoe objective is $\eta_1=1,\eta_2=1,\eta_3=4.3$.}}
	\centering
	\begin{tabular}{|c| c|}
		\hline
		Order  &  Average QoE \\
		\hline
		Z-scan & 11.78  \\
		\hline
		Low $\rightarrow$ High &  10.92 \\
		\hline
		Random & 11.37 \\ 		
		\hline
		High $\rightarrow$ Low & 12.40 \\
		\hline
	\end{tabular}
	\label{order}		
\end{table}
\begin{table}[htbp]
	\caption{\textcolor{black}{the average runtime of 360srl on cpu and gpu platform.}}
	\centering
	\begin{tabular}{|c|  c|}
		\hline
		Average Runtime (ms) & 360SRL \\
		\hline
		CPU  & 22 \\
		\hline 
		GPU  & 10 \\
		\hline
	\end{tabular}
	\label{drl:runtime}		
\end{table}

\section{Conclusion}
In this paper, we focus on optimizing two indispensable components in tile-based  panoramic video adaptive streaming, i.e., viewpoint prediction and rate adaptation. For the first optimization problem, we design a cross-user attentive network, named CUAN, where we enhance the performance of viewpoint prediction through cross-user information extracted by an attention mechanism. For the second optimization problem, we propose a sequential RL-based approach, called 360SRL, which transforms the dimension of action space from exponential to linear by introducing a sequential decision structure.  Experimental results demonstrate that proposed CUAN and 360SRL achieve the state-of-the-art performance. 

We also acknowledge the limitations of our current approach and would like to point out several important future directions to make the method more applicable to real world. First, since the current scheme individually optimizes the viewpoint prediction and rate adaptation lacking consideration of the cascading influence of these two modules, a joint optimization framework could further improve user's QoE. Second, our proposed rate adaptation scheme adopts a \textcolor{black}{simple} QoE metric, whereas the real world QoE metric is much more complicated than this. Although some end-to-end learning-based methods \cite{huang2018qarc} have tried to find more efficient metrics reflecting human visual experience in 2D video streaming, these schemes have not been investigated in the tile-based 360-degree video streaming yet.

\section*{Acknowledgment}
This work was supported in part by NSFC under Grant U1908209, 61571413, 61632001 and the National Key Research and Development Program of China 2018AAA0101400.

\ifCLASSOPTIONcaptionsoff
  \newpage
\fi

\bibliographystyle{IEEEtran}
\bibliography{sample-base}

\begin{thebibliography}{10}
\providecommand{\url}[1]{#1}
\csname url@samestyle\endcsname
\providecommand{\newblock}{\relax}
\providecommand{\bibinfo}[2]{#2}
\providecommand{\BIBentrySTDinterwordspacing}{\spaceskip=0pt\relax}
\providecommand{\BIBentryALTinterwordstretchfactor}{4}
\providecommand{\BIBentryALTinterwordspacing}{\spaceskip=\fontdimen2\font plus
\BIBentryALTinterwordstretchfactor\fontdimen3\font minus
  \fontdimen4\font\relax}
\providecommand{\BIBforeignlanguage}[2]{{%
\expandafter\ifx\csname l@#1\endcsname\relax
\typeout{** WARNING: IEEEtran.bst: No hyphenation pattern has been}%
\typeout{** loaded for the language `#1'. Using the pattern for}%
\typeout{** the default language instead.}%
\else
\language=\csname l@#1\endcsname
\fi
#2}}
\providecommand{\BIBdecl}{\relax}
\BIBdecl

\bibitem{huawei2016}
\BIBentryALTinterwordspacing
Huawei, ``{Whitepaper on the VR-Oriented Bearer Network Requirement},'' Huawei,
  Tech. Rep., 2016. [Online]. Available:
  \url{https://www.huawei.com/en/press-events/news/2016/11/whitepaper-vr-oriented-bearer-network-requirements}
\BIBentrySTDinterwordspacing

\bibitem{youtube2011youtube}
L.~YouTube, ``Youtube,'' \emph{Retrieved}, vol.~27, p. 2011, 2011.

\bibitem{stockhammer2011dynamic}
T.~Stockhammer, ``Dynamic adaptive streaming over http--: standards and design
  principles,'' in \emph{Proceedings of the second annual ACM conference on
  Multimedia systems}.\hskip 1em plus 0.5em minus 0.4em\relax ACM, 2011, pp.
  133--144.

\bibitem{sreedhar2016viewport}
K.~K. Sreedhar, A.~Aminlou, M.~M. Hannuksela, and M.~Gabbouj,
  ``Viewport-adaptive encoding and streaming of 360-degree video for virtual
  reality applications,'' in \emph{2016 IEEE International Symposium on
  Multimedia (ISM)}.\hskip 1em plus 0.5em minus 0.4em\relax IEEE, 2016, pp.
  583--586.

\bibitem{xu2018probabilistic}
Z.~Xu, X.~Zhang, K.~Zhang, and Z.~Guo, ``Probabilistic viewport adaptive
  streaming for 360-degree videos,'' in \emph{2018 IEEE International Symposium
  on Circuits and Systems (ISCAS)}.\hskip 1em plus 0.5em minus 0.4em\relax
  IEEE, 2018, pp. 1--5.

\bibitem{sun2019two}
L.~Sun, F.~Duanmu, Y.~Liu, Y.~Wang, Y.~Ye, H.~Shi, and D.~Dai, ``A two-tier
  system for on-demand streaming of 360 degree video over dynamic networks,''
  \emph{IEEE Journal on Emerging and Selected Topics in Circuits and Systems},
  2019.

\bibitem{sun2018multi}
L.~Sun, F.~Duanmu, Y.~Liu, Y.~Wang, Y.~Ye, and H.~Shi, ``Multi-path multi-tier
  360-degree video streaming in 5g networks,'' in \emph{Proceedings of the 9th
  ACM Multimedia Systems Conference}.\hskip 1em plus 0.5em minus 0.4em\relax
  ACM, 2018, pp. 162--173.

\bibitem{hosseini2016adaptive}
M.~Hosseini and V.~Swaminathan, ``Adaptive 360 vr video streaming: Divide and
  conquer,'' in \emph{2016 IEEE International Symposium on Multimedia
  (ISM)}.\hskip 1em plus 0.5em minus 0.4em\relax IEEE, 2016, pp. 107--110.

\bibitem{xie2017360probdash}
L.~Xie, Z.~Xu, Y.~Ban, X.~Zhang, and Z.~Guo, ``360probdash: Improving qoe of
  360 video streaming using tile-based http adaptive streaming,'' in
  \emph{Proceedings of the 25th ACM international conference on
  Multimedia}.\hskip 1em plus 0.5em minus 0.4em\relax ACM, 2017, pp. 315--323.

\bibitem{ban2018cub360}
Y.~Ban, L.~Xie, Z.~Xu, X.~Zhang, Z.~Guo, and Y.~Wang, ``Cub360: Exploiting
  cross-users behaviors for viewport prediction in 360 video adaptive
  streaming,'' in \emph{2018 IEEE International Conference on Multimedia and
  Expo (ICME)}.\hskip 1em plus 0.5em minus 0.4em\relax IEEE, 2018, pp. 1--6.

\bibitem{le2016tiled}
J.~Le~Feuvre and C.~Concolato, ``Tiled-based adaptive streaming using
  mpeg-dash,'' in \emph{Proceedings of the 7th International Conference on
  Multimedia Systems}.\hskip 1em plus 0.5em minus 0.4em\relax ACM, 2016, p.~41.

\bibitem{xie2018cls}
L.~Xie, X.~Zhang, and Z.~Guo, ``Cls: A cross-user learning based system for
  improving qoe in 360-degree video adaptive streaming,'' in \emph{2018 ACM
  Multimedia Conference on Multimedia Conference}.\hskip 1em plus 0.5em minus
  0.4em\relax ACM, 2018, pp. 564--572.

\bibitem{ban2017optimal}
Y.~Ban, L.~Xie, Z.~Xu, X.~Zhang, Z.~Guo, and Y.~Hu, ``An optimal
  spatial-temporal smoothness approach for tile-based 360-degree video
  streaming,'' in \emph{2017 IEEE Visual Communications and Image Processing
  (VCIP)}.\hskip 1em plus 0.5em minus 0.4em\relax IEEE, 2017, pp. 1--4.

\bibitem{qian2016optimizing}
F.~Qian, L.~Ji, B.~Han, and V.~Gopalakrishnan, ``Optimizing 360 video delivery
  over cellular networks,'' in \emph{Proceedings of the 5th Workshop on All
  Things Cellular: Operations, Applications and Challenges}.\hskip 1em plus
  0.5em minus 0.4em\relax ACM, 2016, pp. 1--6.

\bibitem{xu2018gaze}
Y.~Xu, Y.~Dong, J.~Wu, Z.~Sun, Z.~Shi, J.~Yu, and S.~Gao, ``Gaze prediction in
  dynamic 360 immersive videos,'' in \emph{Proceedings of the IEEE Conference
  on Computer Vision and Pattern Recognition}, 2018, pp. 5333--5342.

\bibitem{fan2017fixation}
C.-L. Fan, J.~Lee, W.-C. Lo, C.-Y. Huang, K.-T. Chen, and C.-H. Hsu, ``Fixation
  prediction for 360 video streaming in head-mounted virtual reality,'' in
  \emph{Proceedings of the 27th Workshop on Network and Operating Systems
  Support for Digital Audio and Video}.\hskip 1em plus 0.5em minus 0.4em\relax
  ACM, 2017, pp. 67--72.

\bibitem{mao2017neural}
H.~Mao, R.~Netravali, and M.~Alizadeh, ``Neural adaptive video streaming with
  pensieve,'' in \emph{Proceedings of the Conference of the ACM Special
  Interest Group on Data Communication}.\hskip 1em plus 0.5em minus 0.4em\relax
  ACM, 2017, pp. 197--210.

\bibitem{gadaleta2017d}
M.~Gadaleta, F.~Chiariotti, M.~Rossi, and A.~Zanella, ``D-dash: A deep
  q-learning framework for dash video streaming,'' \emph{IEEE Transactions on
  Cognitive Communications and Networking}, vol.~3, no.~4, pp. 703--718, 2017.

\bibitem{fu2019360srl}
J.~Fu, X.~Chen, Z.~Zhang, S.~Wu, and Z.~Chen, ``360srl: A sequential
  reinforcement learning approach for abr tile-based 360 video streaming,'' in
  \emph{2019 IEEE International Conference on Multimedia and Expo
  (ICME)}.\hskip 1em plus 0.5em minus 0.4em\relax IEEE, 2019, pp. 290--295.

\bibitem{petrangeli2017http}
S.~Petrangeli, V.~Swaminathan, M.~Hosseini, and F.~De~Turck, ``An http/2-based
  adaptive streaming framework for 360 virtual reality videos,'' in
  \emph{Proceedings of the 25th ACM international conference on
  Multimedia}.\hskip 1em plus 0.5em minus 0.4em\relax ACM, 2017, pp. 306--314.

\bibitem{ester1996density}
M.~Ester, H.-P. Kriegel, J.~Sander, X.~Xu \emph{et~al.}, ``A density-based
  algorithm for discovering clusters in large spatial databases with noise.''
  in \emph{Kdd}, vol.~96, no.~34, 1996, pp. 226--231.

\bibitem{chang2011libsvm}
C.-C. Chang and C.-J. Lin, ``Libsvm: A library for support vector machines,''
  \emph{ACM transactions on intelligent systems and technology (TIST)}, vol.~2,
  no.~3, pp. 1--27, 2011.

\bibitem{sinha1979multiple}
P.~Sinha and A.~A. Zoltners, ``The multiple-choice knapsack problem,''
  \emph{Operations Research}, vol.~27, no.~3, pp. 503--515, 1979.

\bibitem{jiang2014improving}
J.~Jiang, V.~Sekar, and H.~Zhang, ``Improving fairness, efficiency, and
  stability in http-based adaptive video streaming with festive,''
  \emph{IEEE/ACM Transactions on Networking (ToN)}, vol.~22, no.~1, pp.
  326--340, 2014.

\bibitem{sun2016cs2p}
Y.~Sun, X.~Yin, J.~Jiang, V.~Sekar, F.~Lin, N.~Wang, T.~Liu, and B.~Sinopoli,
  ``Cs2p: Improving video bitrate selection and adaptation with data-driven
  throughput prediction,'' in \emph{Proceedings of the 2016 ACM SIGCOMM
  Conference}.\hskip 1em plus 0.5em minus 0.4em\relax ACM, 2016, pp. 272--285.

\bibitem{huang2014buffer}
T.-Y. Huang, R.~Johari, N.~McKeown, M.~Trunnell, and M.~Watson, ``A
  buffer-based approach to rate adaptation: Evidence from a large video
  streaming service,'' in \emph{ACM SIGCOMM Computer Communication Review},
  vol.~44, no.~4.\hskip 1em plus 0.5em minus 0.4em\relax ACM, 2014, pp.
  187--198.

\bibitem{spiteri2016bola}
K.~Spiteri, R.~Urgaonkar, and R.~K. Sitaraman, ``Bola: Near-optimal bitrate
  adaptation for online videos,'' in \emph{IEEE INFOCOM 2016-The 35th Annual
  IEEE International Conference on Computer Communications}.\hskip 1em plus
  0.5em minus 0.4em\relax IEEE, 2016, pp. 1--9.

\bibitem{zhang2019drl360}
Y.~Zhang, P.~Zhao, K.~Bian, Y.~Liu, L.~Song, and X.~Li, ``Drl360: 360-degree
  video streaming with deep reinforcement learning,'' in \emph{IEEE INFOCOM
  2019-IEEE Conference on Computer Communications}.\hskip 1em plus 0.5em minus
  0.4em\relax IEEE, 2019, pp. 1252--1260.

\bibitem{li2019very}
C.~Li, W.~Zhang, Y.~Liu, and Y.~Wang, ``Very long term field of view prediction
  for 360-degree video streaming,'' \emph{arXiv preprint arXiv:1902.01439},
  2019.

\bibitem{cleeremans1989finite}
A.~Cleeremans, D.~Servan-Schreiber, and J.~L. McClelland, ``Finite state
  automata and simple recurrent networks,'' \emph{Neural computation}, vol.~1,
  no.~3, pp. 372--381, 1989.

\bibitem{gers2000recurrent}
F.~A. Gers and J.~Schmidhuber, ``Recurrent nets that time and count,'' in
  \emph{Proceedings of the IEEE-INNS-ENNS International Joint Conference on
  Neural Networks. IJCNN 2000. Neural Computing: New Challenges and
  Perspectives for the New Millennium}, vol.~3.\hskip 1em plus 0.5em minus
  0.4em\relax IEEE, 2000, pp. 189--194.

\bibitem{paszke2017pytorch}
A.~Paszke, S.~Gross, S.~Chintala, and G.~Chanan, ``Pytorch,'' 2017.

\bibitem{kingma2014adam}
D.~P. Kingma and J.~Ba, ``Adam: A method for stochastic optimization,''
  \emph{arXiv preprint arXiv:1412.6980}, 2014.

\bibitem{qian2016qoe}
L.~Qian, Z.~Cheng, Z.~Fang, L.~Ding, F.~Yang, and W.~Huang, ``A qoe-driven
  encoder adaptation scheme for multi-user video streaming in wireless
  networks,'' \emph{IEEE Transactions on Broadcasting}, vol.~63, no.~1, pp.
  20--31, 2016.

\bibitem{yu2017qoe}
L.~Yu, T.~Tillo, and J.~Xiao, ``Qoe-driven dynamic adaptive video streaming
  strategy with future information,'' \emph{IEEE Transactions on Broadcasting},
  vol.~63, no.~3, pp. 523--534, 2017.

\bibitem{doumanoglou2018quality}
A.~Doumanoglou, D.~Griffin, J.~Serrano, N.~Zioulis, T.~K. Phan, D.~Jim{\'e}nez,
  D.~Zarpalas, F.~Alvarez, M.~Rio, and P.~Daras, ``Quality of experience for
  3-d immersive media streaming,'' \emph{IEEE Transactions on Broadcasting},
  vol.~64, no.~2, pp. 379--391, 2018.

\bibitem{tam2011stereoscopic}
W.~J. Tam, F.~Speranza, S.~Yano, K.~Shimono, and H.~Ono, ``Stereoscopic 3d-tv:
  visual comfort,'' \emph{IEEE Transactions on Broadcasting}, vol.~57, no.~2,
  pp. 335--346, 2011.

\bibitem{chen2015multimedia}
C.~W. Chen, P.~Chatzimisios, T.~Dagiuklas, and L.~Atzori, \emph{Multimedia
  quality of experience (QoE): current status and future requirements}.\hskip
  1em plus 0.5em minus 0.4em\relax John Wiley \& Sons, 2015.

\bibitem{de2012quantifying}
T.~De~Pessemier, K.~De~Moor, W.~Joseph, L.~De~Marez, and L.~Martens,
  ``Quantifying the influence of rebuffering interruptions on the user's
  quality of experience during mobile video watching,'' \emph{IEEE Transactions
  on Broadcasting}, vol.~59, no.~1, pp. 47--61, 2012.

\bibitem{singh1996make}
S.~Singh, P.~Norvig, D.~Cohn \emph{et~al.}, ``How to make software agents do
  the right thing: An introduction to reinforcement learning,'' \emph{Adaptive
  Systems Group}, 1996.

\bibitem{mnih2016asynchronous}
V.~Mnih, A.~P. Badia, M.~Mirza, A.~Graves, T.~Lillicrap, T.~Harley, D.~Silver,
  and K.~Kavukcuoglu, ``Asynchronous methods for deep reinforcement learning,''
  in \emph{International conference on machine learning}, 2016, pp. 1928--1937.

\bibitem{sutton1998reinforcement}
R.~S. Sutton and A.~G. Barto, ``Reinforcement learning: an introduction mit
  press,'' \emph{Cambridge, MA}, 1998.

\bibitem{merritt2006x264}
L.~Merritt and R.~Vanam, ``x264: A high performance h. 264/avc encoder,''
  \emph{online] http://neuron2. net/library/avc/overview\_x264\_v8\_5. pdf},
  2006.

\bibitem{riiser2013commute}
H.~Riiser, P.~Vigmostad, C.~Griwodz, and P.~Halvorsen, ``Commute path bandwidth
  traces from 3g networks: analysis and applications,'' in \emph{Proceedings of
  the 4th ACM Multimedia Systems Conference}, 2013, pp. 114--118.

\bibitem{huang2015buffer}
T.-Y. Huang, R.~Johari, N.~McKeown, M.~Trunnell, and M.~Watson, ``A
  buffer-based approach to rate adaptation: Evidence from a large video
  streaming service,'' \emph{ACM SIGCOMM Computer Communication Review},
  vol.~44, no.~4, pp. 187--198, 2015.

\bibitem{huang2018qarc}
T.~Huang, R.-X. Zhang, C.~Zhou, and L.~Sun, ``Qarc: Video quality aware rate
  control for real-time video streaming based on deep reinforcement learning,''
  in \emph{Proceedings of the 26th ACM international conference on Multimedia},
  2018, pp. 1208--1216.

\end{thebibliography}
\begin{IEEEbiography}[{\includegraphics[width=1in,height=1.25in,clip,keepaspectratio]{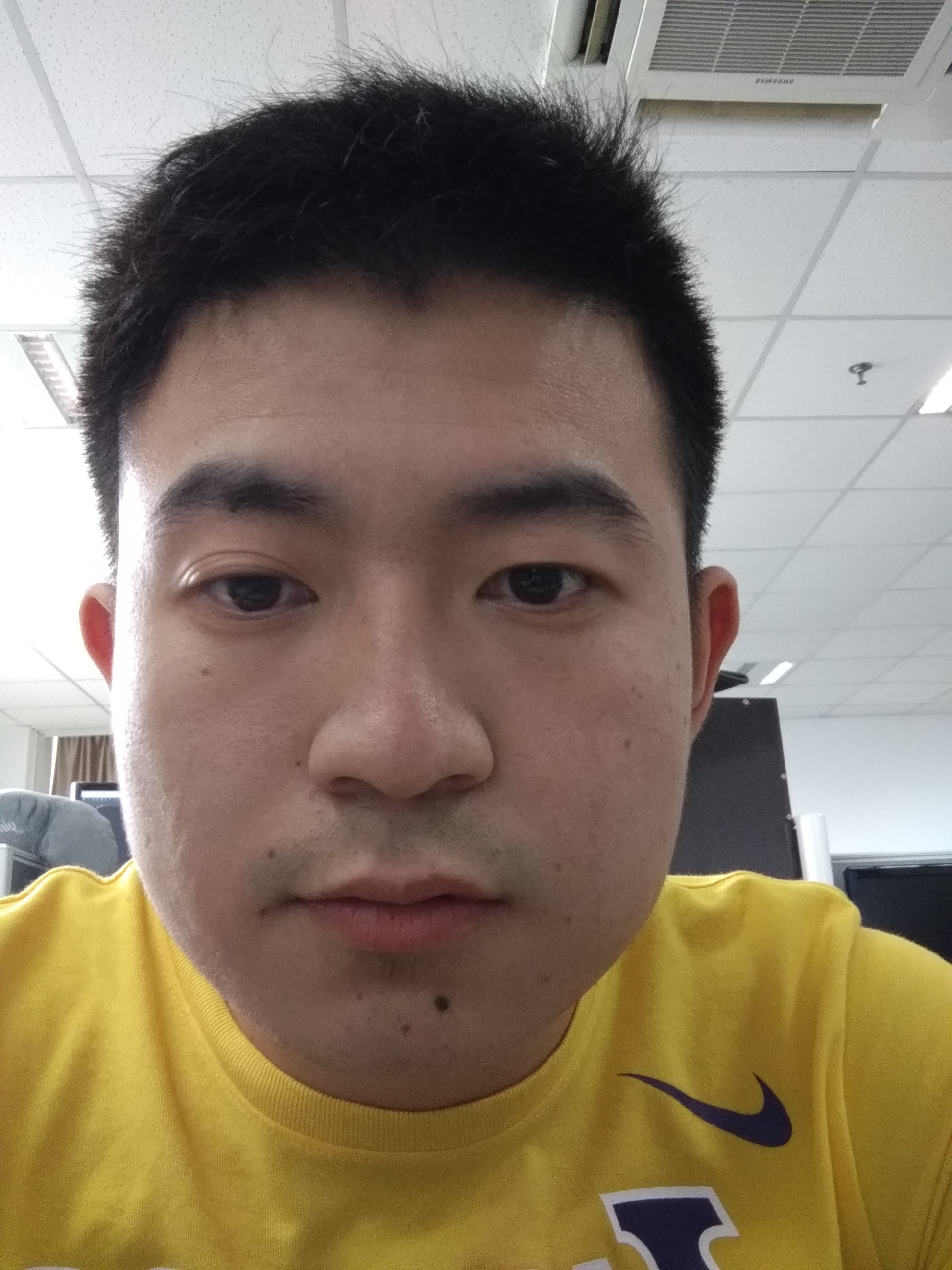}}]{Jun Fu} received the B.S. degree in Department of Electrical Engineering from Chongqing University in 2017, and is currently studying for Ph.D. degree at University of Science and Technology of China (USTC), Hefei, China.  His research interests include immersive media streaming, image and video compression, and automatic machine learning. 
\end{IEEEbiography}
\begin{IEEEbiography}[{\includegraphics[width=1in,height=1.25in,clip,keepaspectratio]{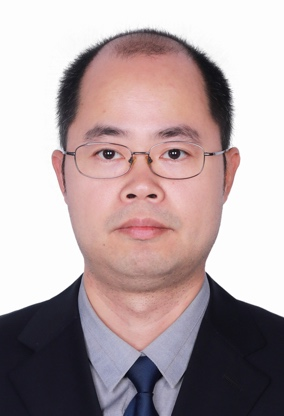}}]{Zhibo Chen} received the B. Sc., and Ph.D. degree from Department of Electrical Engineering Tsinghua University in 1998 and 2003, respectively. He is now a professor in University of Science and Technology of China. His research interests include image and video compression, visual quality of experience assessment, immersive media computing and intelligent media computing. He has more than 100 publications and more than 50 granted EU and US patent applications. He is IEEE senior member, Secretary of IEEE Visual Signal Processing and Communications Committee, and member of IEEE Multimedia System and Applications Committee. He was TPC chair of IEEE PCS 2019 and organization committee member of ICIP 2017 and ICME 2013, served as TPC member in IEEE ISCAS and IEEE VCIP.
\end{IEEEbiography}
\begin{IEEEbiography}[{\includegraphics[width=1in,height=1.25in,clip,keepaspectratio]{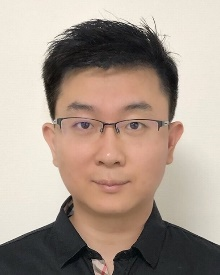}}]{Xiaoming Chen} received the B.Sc. degree from the Royal Melbourne Institute of Technology, Australia, and the Ph.D. degree from The University of Sydney, Australia, in 2004 and 2009, respectively. From 2010 to 2014, he was with the National University of Singapore, CSIRO Australia, and IBM. From 2014 to 2019, he had been a Researcher with the Institute of Advanced Technology, University of Science and Technology of China. In 2014, he was invited into the “100-Talent Program” by the government of Hefei, Anhui Province, China. He is currently a Professor with the School of Computer Science and Engineering, Beijing Technology and Business University, China. Meanwhile, he has been invited as a Guest Researcher with Beijing Research Institute, University of Science and Technology of China since 2019. His research interests include immersive media computing, virtual reality, and related business information systems. His work has been published in ACM MM, IEEE Virtual Reality, IEEE TIP, IEEE T-CSVT, etc. 

\end{IEEEbiography}
\begin{IEEEbiography}[{\includegraphics[width=1in,height=1.25in,clip,keepaspectratio]{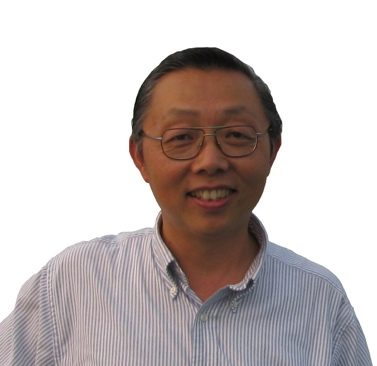}}]{Weiping Li} received the B.S. degree in electrical engineering from University of Science and Technology of China (USTC), Hefei, China, in 1982, and the M.S. and Ph.D. degrees in electrical engineering from Stanford University, Stanford, CA, USA, in 1983 and 1988, respectively. In 1987, he joined Lehigh University, Bethlehem, PA, USA, as an Assistant Professor with the Department of Electrical Engineering and Computer Science. In 1993, he was promoted to Associate Professor with tenure. In 1998, he was promoted to Full Professor. From 1998 to 2010, he worked in several high-tech companies in the Silicon Valley (1998-2000, Optivision, Palo Alto; 2000-2002, Webcast Technologies, Mountain View; 2002-2008, Amity Systems, Milpitas, 2008-2010, Bada Networks, Santa Clara; all in California, USA). In March 2010, he returned to USTC to serve as the Dean of the School of Information Science and Technology until July 2014 and is currently a Professor with the School of Information Science and Technology. Dr. Li had served as the Editor-in-Chief of IEEE TRANSACTIONS ON CIRCUITS AND SYSTEMS FOR VIDEO TECHNOLOGY and Guest Editor of the PROCEEDINGS OF THE IEEE. He was the Chair of several Technical Committees in the IEEE Circuits and Systems Society and IEEE International Conferences, and the Chair of the Best Student Paper Award Committee for SPIE Visual Communications and Image Processing Conference. He has made many contributions to international standards. His inventions on fine granularity scalable video coding and shape adaptive wavelet coding have been included in the MPEG-4 international standard. He served as a Member of the Moving Picture Experts Group (MPEG) of the International Standard Organization (ISO) and an Editor of MPEG-4 international standard. He served as a Founding Member of the Board of Directors of MPEG-4 Industry Forum. As a Technical Advisor, he also made contributions to the Chinese audio video coding standard and its applications. He was the recipient of the Certificate of Appreciation from ISO/IEC as a Project Editor in development of an international standard in 2004, the Spira Award for Excellence in Teaching in 1992 at Lehigh University, and the first Guo Mo-Ruo Prize for Outstanding Student in 1980 at USTC.
\end{IEEEbiography}
\end{document}